\documentclass[preprint,prd,aps,showpacs,preprintnumbers,amsmath,amssymb]{revtex4-1}
\usepackage{graphics,epsfig,subfigure}
\usepackage{diagbox}
\usepackage[usenames]{color}
\usepackage{color}
\usepackage{graphicx}
\usepackage{amsfonts}
\usepackage{indentfirst}
\usepackage{booktabs}
\usepackage{hyperref}
\hypersetup{hypertex=ture,backref=true,colorlinks=ture,linkcolor=blue,anchorcolor=blue,citecolor=blue}
\usepackage{float}
\usepackage{multirow}
\usepackage[section]{placeins}
\usepackage{array}
\usepackage{amsmath}
\usepackage{slashed}
\usepackage{caption}
\usepackage{fancyhdr}

\begin{document}
\title{Bardeen-Dirac Stars in AdS Spacetime}
\author{Xiao-Yu Zhang}
\author{Li Zhao\footnote{ lizhao@lzu.edu.cn, corresponding author}}
\author{Yong-Qiang Wang\footnote{yqwang@lzu.edu.cn, corresponding author}}

\affiliation{ $^{1}$Lanzhou Center for Theoretical Physics, Key Laboratory of Theoretical Physics of Gansu Province,School of Physical Science and Technology, Lanzhou University, Lanzhou 730000, China\\
$^{2}$Institute of Theoretical Physics $\&$ Research Center of Gravitation, Lanzhou University, Lanzhou 730000, China}

\begin{abstract}
In this paper, we construct a static spherical symmetric Bardeen-Dirac Stars (BDSs) in the four-dimensional Anti-de Sitter (AdS) spacetime, which consists of the electromagnetic field and Dirac field coupled to gravity. We investigate the ADM mass, Noether charge and light rings of BDSs in AdS spacetime. In asymptotically Minkowski spacetime, the maximum frequency of BDSs is one. However, we observe that the maximum frequency of BDSs increases as the cosmological constant decreases in AdS spacetime. Additionally, BDSs can exhibit extreme behavior at low frequencies, refer to as Frozen Bardeen-Dirac stars (FBDSs) in AdS spacetime. FBDSs have a critical event horizon, where the metric function $g_{tt}$ is very close to zero. The matter is entirely encapsulated by this critical horizon, highly concentrated within it. When the magnetic charge is fixed, the FBDSs gradually disappear as the cosmological constant decreases.
\end{abstract}

\maketitle
\thispagestyle{empty}

\section{INTRODUCTION}\label{Sec1}
In 1916, Einstein introduced General Relativity (GR) \cite{Einstein:1916vd}, an extension and refinement of Newtonian gravity that significantly deepened our understanding of gravitational phenomena. GR provides a more accurate description under more extreme conditions, especially in strong gravitational fields. Although GR has greatly expanded our understanding of gravity, the problem of singularities remains unresolved. Also in 1916, Schwarzschild provided the first black hole solution based on Einstein's equations of general relativity \cite{Schwarzschild:1916uq}, known as the Schwarzschild black hole. There is a singularity at the center of a Schwarzschild black hole where the curvature of spacetime becomes infinite. The Schwarzschild solution was the first to reveal the presence of singularities inside black holes. As research progressed, more complex black hole solutions such as the Kerr black hole (describing rotating black holes) \cite{Kerr:1963ud} and the Kerr–Newman black hole (describing charged and rotating black holes) \cite{Newman:1965my} also revealed the presence of singularities, though they have different forms. Singularities are not a reflection of the real physical world but rather a limitation of existing theories when describing the behavior of matter and spacetime under extreme conditions.

Therefore, addressing the singularity problem has been a major challenge in gravitational research. To avoid the occurrence of singularities, regular black holes are one of many theoretical models proposed to address the issue of black hole singularities. These models modify traditional gravitational theories or introduce new physical mechanisms to prevent the formation of singularities with infinite curvature at the black hole's center, offering a new way to describe the spacetime structure under extreme gravitational conditions. The idea of regular black holes originated with Sakharov \cite{Sakharov:1966aja} and Gliner \cite{Gliner:1966} in 1966. In 1968, Bardeen created the Bardeen black hole \cite{Bardeen:1968} which avoids forming a singularity in the traditional sense at the black hole's center by introducing matter field.  Later, Beato and Garcia derived the Bardeen black hole solution \cite{Ayon-Beato:2000mjt} by coupling gravity with a nonlinear electromagnetic field. Since then, research on regular black holes has increasingly expanded \cite{Dymnikova:1992ux,Ayon-Beato:1998hmi,Balart:2014cga,Balart:2016zrd,Rodrigues:2018bdc,deSousaSilva:2018kkt,Balart:2014jia,Bambi:2013ufa,Lan:2020fmn,Bueno:2024dgm,Barenboim:2024dko}. On the other hand, the exotic compact objects are theoretical models constructed with exotic matter that can also avoid the presence of singularities. Among these, boson stars are the most widely studied and are considered one of the potential candidates for black holes \cite{Guzman:2009zz,Cunha:2015yba,Cardoso:2019rvt}. Research on boson stars can be traced back to the pioneering work of Ruffni \cite{Ruffini:1969qy} and Kupa \cite{Kaup:1968zz} in the 1960s. Since then, research on boson stars has become quite extensive \cite{Li:2020ffy,Herdeiro:2020jzx,Sun:2022duv,Zhang:2021xhp,Kleihaus:2009kr}. By further extending the scalar field to vector fields and spinor fields, one can construct Proca stars \cite{Brito:2015pxa,SalazarLandea:2016bys,Su:2023zhh,Zhang:2023rwc} and Dirac stars \cite{Finster:1998ws,Finster:1998ux,Dzhunushaliev:2018jhj,Huang:2023glq,Hao:2023igi}, which further expand the theoretical framework of the exotic compact objects.

Considering both Bardeen's theory and matter field simultaneously leads to interesting phenomena in this new theoretical model. For example, introducing the scalar field into the Bardeen's theory \cite{Wang:2023tdz}, resulting in Bardeen-boson stars (BBSs). The introduction of the scalar field restricts the magnetic charge, which can no longer take arbitrary values but must be limited to a critical range. Additionally, when the magnetic charge is constrained to a smaller range, it leads to more distinct properties. The frequency of the scalar field can approach zero. Low-frequency BBSs possess a critical horizon, which differs from the event horizon of a black hole. At this point, the component of the metric $g_{tt}$ can be very close to zero, reaching values as small as $10^{-5}$ or even smaller. The scalar matter is confined within the critical horizon. Due to the similarity in properties between this solution and those of frozen stars, low-frequency BBSs are also referred to as Frozen Bardeen-Boson Stars (FBBSs). In addition, there are studies on charged BBSs \cite{Huang:2024rbg} and Hayward axion stars \cite{Chen:2024bfj} when considering scalar fields. Extending the scalar field to a Dirac field resulting in Bardeen-Dirac stars (BDSs) \cite{Huang:2023fnt}. Although the matter field employs the Dirac field, similar properties and frozen star solutions (FBDSs) also exist. Inspired by the AdS/CFT correspondence \cite{Maldacena:1997re,Witten:1998qj}, research on these theoretical models in AdS spacetime has become quite extensive. For example, research on the regular black holes \cite{Xie:2024xkh,Guo:2024jhg,Fan:2016rih}, boson stars \cite{Astefanesei:2003qy,Hartmann:2012gw,Brihaye:2014bqa,Brihaye:2022oaf,Guo:2020bqz,Kichakova:2013sza},  as well as their extensions, including Proca stars \cite{Duarte:2016lig} and Dirac stars \cite{Zhang:2024nju}. Therefore, we are very interested in studying the effect of the cosmological constant on BDSs. By introducing a negative cosmological constant, we can construct BDSs in AdS spacetime. The study primarily focuses on the ADM mass and Noether charge of BDSs.  Furthermore, we extend our research to the light rings of BDSs. Similar to FBDSs in asymptotically Minkowski spacetime, the inner light ring becomes a light ball.

The structure of this paper is as follows. In Sec. \ref{sec2}, we construct Bardeen-Dirac Stars in the four-dimensional AdS spacetime, consisting of gravity coupled to an electromagnetic field and two Dirac fields. In Sec. \ref{sec3}, we present the boundary conditions. Sec. \ref{sec4} displays the numerical results of the equations of motion and analyzes the properties of BDSs. Finally, we summarize our findings and discuss the numerical results in Sec. \ref{sec5}.
\section{THE MODEL SETUP}\label{sec2}
\subsection{Framework}
We consider the system of Einstein gravity to be minimally coupled to two Dirac fields and the nonlinear
electromagnetic field of the Bardeen model in 3+1-dimensional spacetime, and the corresponding action of the system is
    \begin{equation}\label{action}
        S=\int d^4x\sqrt{-g}\left(\frac{R-2\Lambda}{16\pi G}+\mathcal{L}^{D}+\mathcal{L}^{B}\right),
    \end{equation}
where $G$ and $\Lambda$ are the Newtonian constant and the cosmological constant, respectively. $R$ is the Ricci scalar. The two Lagrangian densities in the action correspond to the Dirac field and the electromagnetic field, respectively. Their forms are given by
    \begin{eqnarray}
        \mathcal{L}^{D}=\sum_{k=1}^{2}\mathcal{L}^{D (k)} = -\sum_{k=1}^{2} \left[\frac{\mathrm{i}}{2}\left(\left\{\hat{\slashed{D}} \overline{\Psi}^{(k)}\right\}\Psi^{(k)} - \overline{\Psi}^{(k)} \hat{\slashed{D}} \Psi^{(k)} \right) + \mu\overline{\Psi}^{(k)}\Psi^{(k)} \right],
    \end{eqnarray}
    \begin{eqnarray}
        &&\mathcal{L}^{B}= -\frac{3}{2s}\left(\frac{\sqrt{2q^{2}\mathcal{F}}}{1+\sqrt{2q^{2}\mathcal{F}}}\right)^{5/2}.
    \end{eqnarray}
Here the two Dirac fields have spherically symmetric configuration, $\overline{\Psi}^{(k)}$ are the Dirac conjugate, and $\mu$ is the mass of the Dirac field. The operator $\hat{\slashed{D}}=\gamma^{\mu}\hat{D}_{\mu}$ involves the gamma matrices $\gamma^{\mu}$ in curved spacetime. $\hat{D}_{\mu}=\partial_{\mu}-\Gamma_{\mu}$ is the spinor covariant derivative, where $\Gamma_{\mu}$ is the spinor connection matrices. The term $\mathcal{F}=\frac{1}{4}F_{\mu\nu}F^{\mu\nu}$ is defined with the electromagnetic field $F_{\mu\nu}=\partial_{\mu}A_{\nu}-\partial_{\nu}A_{\mu}$, where $A_{\mu}$ is the electromagnetic 4-potential. $q$ and $s$ are constants, with $q$ representing the magnetic charge. The field equations are given by
    \begin{eqnarray}
        &&\mathcal{R}_{\alpha\beta}-\frac{1}{2}\mathcal{R} \mathrm{g}_{\alpha\beta}+\Lambda \mathrm{g}_{\beta}=8 \pi G \left(T^{D}_{\alpha\beta}+ T^{B}_{\alpha\beta}\right),
    \end{eqnarray}
    \begin{eqnarray}
        \left(\hat{\slashed{D}}-\mu\right)\Psi^{(k)}=0,
    \end{eqnarray}
    \begin{eqnarray}
         \nabla_{\alpha}\left(\frac{\partial\mathcal{L}^{B}}{\partial\mathcal{F}}F^{\alpha}_{\beta}\right)=0,
    \end{eqnarray}
where $T^{D}_{\alpha\beta}$ and $T^{B}_{\alpha\beta}$ are the energy-momentum tensors of the Dirac field and  the electromagnetic field, respectively. They are given by
    \begin{eqnarray}
        T^{D}_{\alpha\beta}=\sum_{k=1}^{2}\left[-\frac{\mathrm{i}}{2}\left(\overline{\Psi}^{(k)}\gamma_{\left(\alpha\right.}\hat{D}_{\left.\beta\right)}\Psi^{(k)}
        -\left\{\hat{D}_{\left(\alpha\right.}\overline{\Psi}^{(k)} \right\}\gamma_{\left.\beta\right)} \Psi^{(k)} \right)-\mathrm{g}_{\alpha\beta}\mathcal{L}^{D(k)}\right].
    \end{eqnarray}
    \begin{eqnarray}
         T^{B}_{\alpha\beta}=-\frac{\partial\mathcal{L}^{B}}{\partial\mathcal{F}}F_{\alpha\sigma}F_\beta{ }^\sigma +\mathrm{g}_{\alpha\beta}\mathcal{L}^{B}.
    \end{eqnarray}
The action of the matter field is invariant under the $ U(1)$ transformation $\Psi^{(k)}\rightarrow e^{i\alpha}\Psi^{(k)}$, where $\alpha$ is a constant. Therefore, there exists a conserved current
    \begin{eqnarray}
        &&j^{\alpha} = \sum_{k=1}^{2}\overline{\Psi}^{(k)}\gamma^{\alpha}\Psi^{(k)},
    \end{eqnarray}
and the conserved charge can be obtained
    \begin{eqnarray}
        &&Q = \int_{\Sigma} j^{\mathrm{t}},\label{Q}
    \end{eqnarray}
where $ j^{\mathrm{t}}$ is the time component of the current, and $\Sigma$ is a timelike hypersurface.
\subsection{Ansatz and equation of motion}
We consider the following static spherically symmetric metric
    \begin{eqnarray}
        ds^{2}=-N\left(r\right)\sigma^{2}\left(r\right)dt^{2}
        +\frac{dr^2}{N\left(r\right)}
        +r^{2}\left(d \theta^{2}+\sin^{2}\theta d \varphi^{2} \right),
    \end{eqnarray}
where $N(r)=1-\frac{2m(r)}{r}-\frac{\Lambda r^{2}}{3}$, and $m(r)$ depends only on the radial distance $r$. For the Dirac field and  the electromagnetic field, we use the following ansatz
    \begin{equation}
         \Psi^{(1)} = \begin{pmatrix}\cos(\frac{\theta}{2})[(1 + i)f(r) - (1 - i)g(r)]\\ i\sin(\frac{\theta}{2})[(1 - i)f(r) - (1 + i)g(r)]\\-i\cos(\frac{\theta}{2})[(1 - i)f(r) - (1 + i)g(r)]\\ -\sin(\frac{\theta}{2})[(1 + i)f(r) - (1 - i)g(r)] \end{pmatrix}e^{i\frac{\varphi}{2} - i\omega t}\,,
    \end{equation}
    \begin{equation}
         \Psi^{(2)} = \begin{pmatrix}i\sin(\frac{\theta}{2})[(1 + i)f(r) - (1 - i)g(r)]\\ \cos(\frac{\theta}{2})[(1 - i)f(r) - (1 + i)g(r)]\\ \sin(\frac{\theta}{2})[(1 - i)f(r) - (1 + i)g(r)]\\ i\cos(\frac{\theta}{2})[(1 + i)f(r) - (1 - i)g(r)] \end{pmatrix}e^{-i\frac{\varphi}{2} - i\omega t}\,,
    \end{equation}
    \begin{equation}
         A = q \cos\left(\theta\right)d\phi\,,
    \end{equation}
where $f(r)$ and $g(r)$ are real functions, and $\omega$ is the frequency of the Dirac field.
By substituting the ansatz into the Einstein equation and the Dirac equation, we obtain
    \begin{eqnarray}\label{12-15}
        &&m^{\prime} = 32 \pi Gr^{2}\left[\sqrt{N}\left(gf^{\prime}-fg^{\prime}\right)+\frac{2fg}{r}
        + \mu\left(g^{2}-f^{2}\right) \right]+\frac{6 \pi Gq^{5}r^{2}}{s\left(q^{2}+r^{2}\right)^{5/2}},\label{eom15}   \\
        &&\frac{\sigma^{\prime}}{\sigma}= \frac{32\pi G r}{\sqrt{N}}\left(gf^{\prime}-fg^{\prime}
        +\frac{\omega\left(f^{2}+g^{2}\right)}{N\sigma}\right), \label{eom16}     \\
        &&f^{\prime}+\left(\frac{N^{\prime}}{4N}+\frac{\sigma^{\prime}}{2\sigma}+\frac{1}{r\sqrt{N}}+\frac{1}{r}\right)f
        -\frac{\omega g}{N \sigma}+\frac{g \mu}{\sqrt{N}} =0,         \\
        &&g^{\prime}+\left(\frac{N^{\prime}}{4N}+\frac{\sigma^{\prime}}{2\sigma}-\frac{1}{r\sqrt{N}}+\frac{1}{r}\right)f
        +\frac{\omega f}{N \sigma}+\frac{f \mu}{\sqrt{N}} =0.\label{eom18}
    \end{eqnarray}
The equations here are quite similar to those of asymptotically Minkowski spacetime \cite{Huang:2023fnt}, with the only difference being the inclusion of a cosmological constant in the $N(r)$.  The conserved charge is
    \begin{eqnarray}
        Q=8\int_{0}^{\infty}dr \, r^{2}\frac{\left(f^{2}+g^{2}\right)}{\sqrt{N}},
    \end{eqnarray}
and the energy density is $\rho=-g^{0\alpha}T_{\alpha0}^{D}$
    \begin{eqnarray}
        \rho=8\left(\left(gf^{\prime}-fg^{\prime}\right)\sqrt{N}+\frac{2fg}{r}+\mu\left(g^{2}-f^{2}\right)\right).
    \end{eqnarray}
\subsection{The effective potential}\label{c}
In order to study the properties of the light ring, we first need to analyze the effective potential experienced by particles on null paths as they orbit around the BDSs. The expression for the effective potential is derived from the geodesic equation of the photon, given by
    \begin{eqnarray}
       g_{\alpha\beta}\frac{dx^{\alpha}}{d\lambda}\frac{dx^{\beta}}{d\lambda}=0.
    \end{eqnarray}
Here $\lambda$ is the affine parameter of the geodesic. Due to our choice of a static, spherically symmetric configuration, which preserves time translation symmetry and rotational symmetry. There are two conserved quantities: energy $E=-g_{tt}dt/d\lambda$ and angular momentum $L=r^{2}d\varphi/d\lambda$. By substituting the ansatz, we can obtain
    \begin{eqnarray}
       \left(\frac{dr}{d\lambda}\right)^{2}+\frac{L^{2}}{g_{tt}g_{rr}}\left(\frac{1}{b^{2}}+\frac{g_{tt}}{r^{2}}\right)= \left(\frac{dr}{d\lambda}\right)^{2}+\frac{L^{2}}{\sigma^{2}}\left(\frac{N\sigma^{2}}{r^{2}}-\frac{1}{b^{2}}\right)=0,
    \end{eqnarray}
where the impact parameter $b=L/E$ is a constant. We define the effective pitential $V_{eff}$ is
    \begin{eqnarray}
       V_{eff}=\frac{N\sigma^{2}}{r^{2}}.\label{eom23}
    \end{eqnarray}
The position of the light ring $R_{LR}$, is determined by the extremum of the effective potential, so the effective potential should satisfy the following condition
    \begin{eqnarray}
       \left.\frac{dV_{eff}}{dr}\right|_{R_{LR}}=0.
    \end{eqnarray}
In this way, we can determine the position of the light ring. The stability of the light ring is determined by the second derivative of the effective potential at the position of the light ring.  If $V_ {eff}^{''}<0$, the light ring is unstable. Conversely, if the second derivative is greater than zero, the light ring is stable.

\section{BOUNDARY CONDITIONS}\label{sec3}
For spherically symmetric BDSs, when computing the numerical solutions, it is necessary to choose appropriate boundary conditions. The conditions that the metric functions need to satisfy at the origin and at infinity are
    \begin{eqnarray}
        m\left(0 \right)=0, \quad \sigma\left(0 \right)=\sigma_{0}, \quad
        m\left(\infty \right)= M, \quad \sigma\left(\infty \right)=1,
    \end{eqnarray}
where $M$ and $\sigma_0$ are still unknown. $M$ represents the ADM mass. The conditions for the Dirac field at the origin and at infinity are
    \begin{eqnarray}
        f\left(0 \right)=0,  \quad  \left.\frac{dg\left(r \right)}{dr}\right|_{r=0}=0,\quad f\left(\infty \right)=g\left(\infty \right)=0.
    \end{eqnarray}
\section{NUMERICAL RESULTS}\label{sec4}
To simplify the numerical calculations, we need to use dimensionless transformations. The system of equations depends only on the Dirac field mass $\mu$, the gravitational constant $G$, the cosmological constant $\Lambda$, the magnetic charge $q$ and coupling parameter $s$ in the Bardeen action, where $q$ and $s$ are dimensionless quantities. In this case, we adopt
\begin{eqnarray}
        r\rightarrow r\mu,\quad\omega\rightarrow\omega/\mu,\quad \nu\rightarrow\nu \mu
        , \quad \Lambda\rightarrow \Lambda/\mu^{2},\quad f\rightarrow \sqrt{\frac{4 \pi G}{\mu}}f ,\quad g\rightarrow \sqrt{\frac{4 \pi G}{\mu}}g.
    \end{eqnarray}
We set the units $\mu=1$ and $4 \pi G=1$, which is a common approach when studying boson stars. Additionally, we set $s=0.5$ throughout the paper. After setting the units, our input parameters are only $\omega$, $q$ and $\Lambda$. We also need to introduce a new radial transformation
     \begin{eqnarray}
         x=\frac{r}{1+r}.
     \end{eqnarray}
Here the radial coordinate is $r\in[0,\infty)$, so $x\in[0,1]$. We use the Newton-Raphson method as the iterative technique and apply the finite element method for numerical solutions of the differential equations. To ensure the accuracy of the solutions, we require the relative error to be less than $10^{-5}$.

Before discussing our numerical results, it is important to mention two special cases. For $q=0$, our model reduces to the Einstein-Dirac model, which corresponds to a spherically symmetric Dirac stars. Another case is when $q\neq0$ and the Dirac field vanishes, which describes a spherically symmetric Bardeen black hole. In asymptotically Minkowski spacetime, from the metric of the Bardeen black hole, we find that the magnetic charge must satisfy $q < 3^{3/4}\sqrt{s/2}$. This condition ensures that no event horizon appears for the Bardeen black hole. Specifically, for $q \geq 3^{3/4}\sqrt{s/2}$, we cannot find BDSs, which is also evident from the subsequent numerical results. Introducing a negative cosmological constant will alter this range. However, we cannot obtain a specific analytical result. In our paper, when $\Lambda=-0.1,s=0.5$, this maximum value is approximately $1.867$. Therefore, the magnetic charge is restricted to $q<1.867$ in our subsequent discussions.

         \begin{figure}
            \centering
            \subfigure{\includegraphics[width=0.45\textwidth]{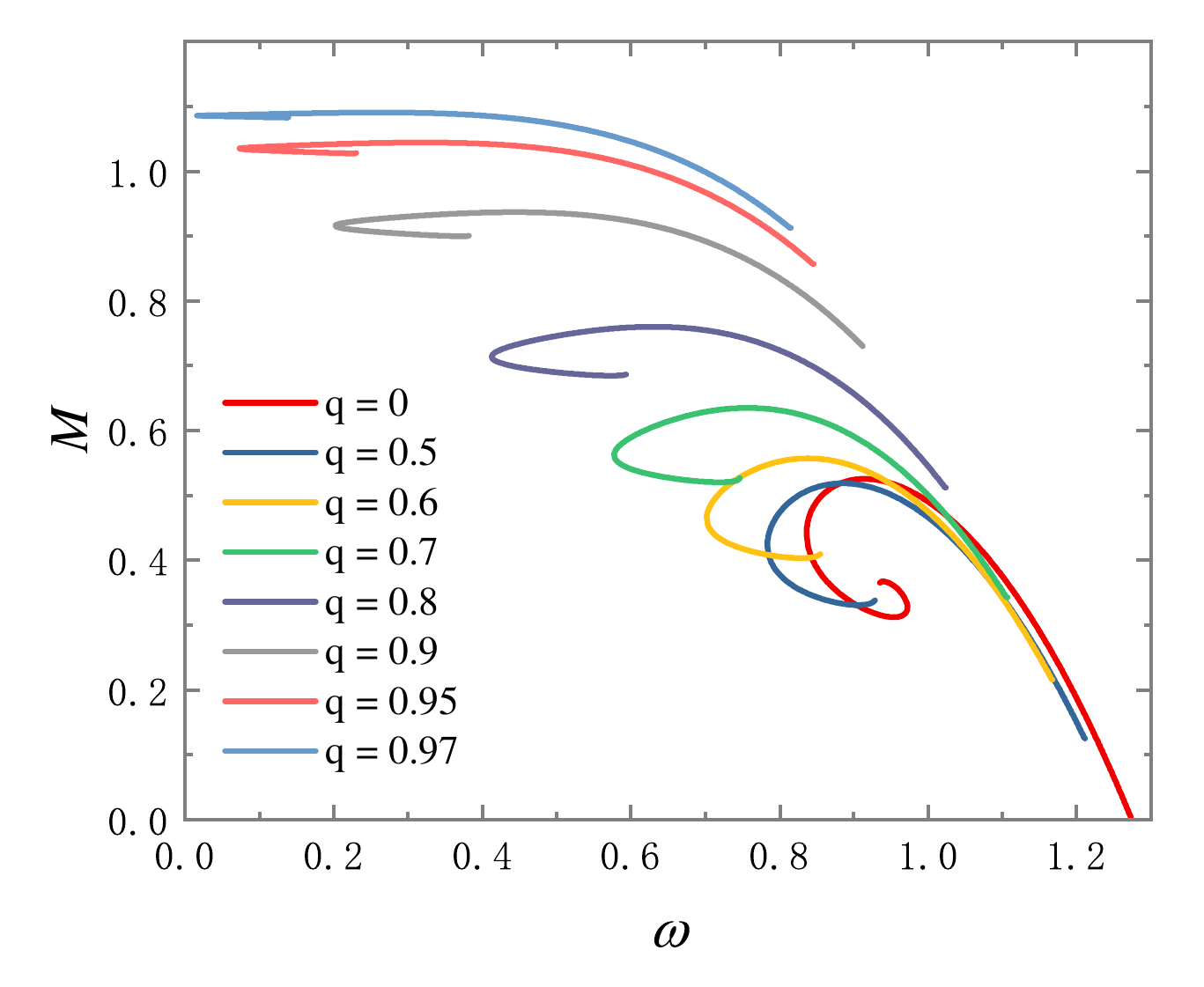}}
            \subfigure{\includegraphics[width=0.45\textwidth]{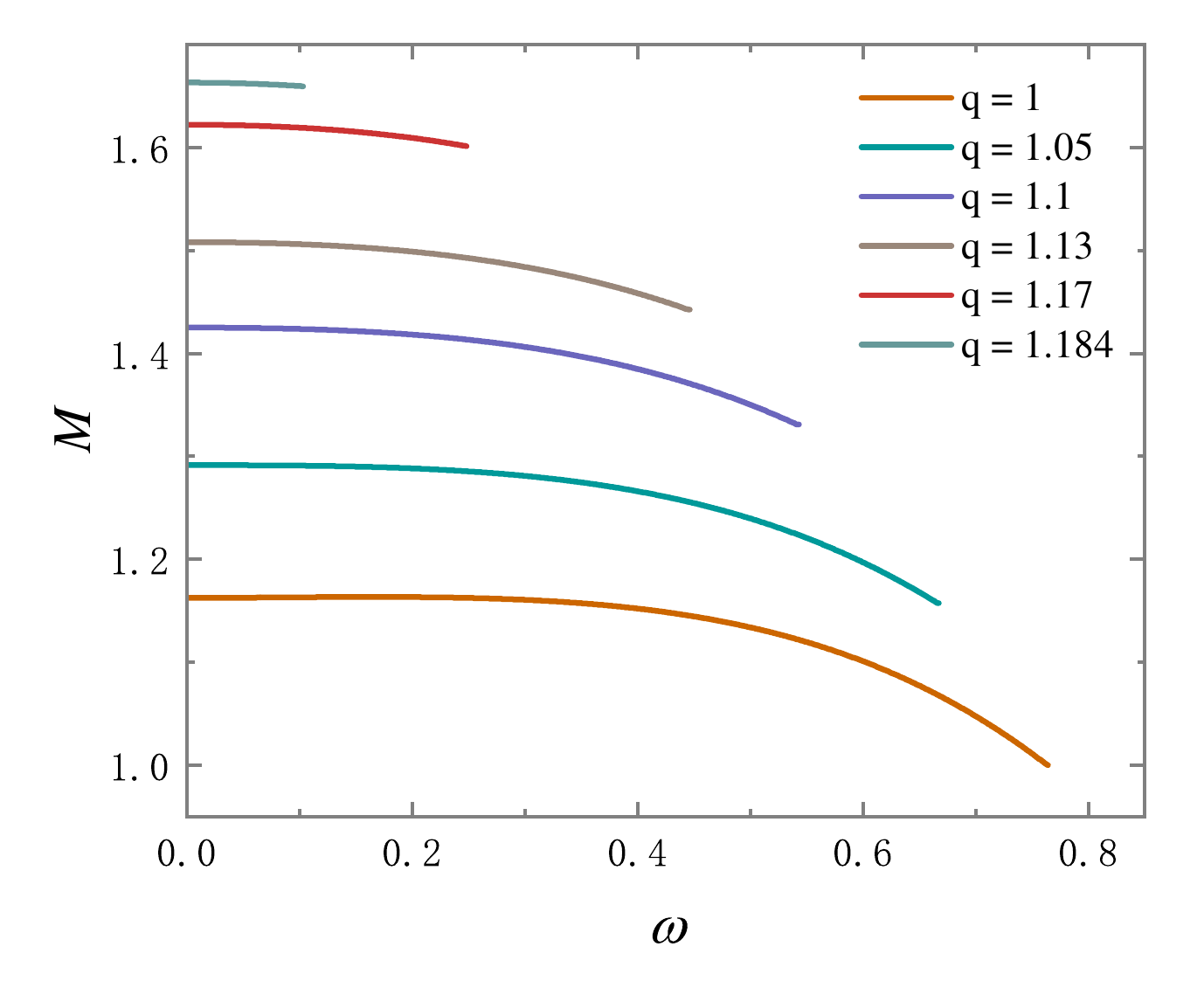}}
            \subfigure{\includegraphics[width=0.45\textwidth]{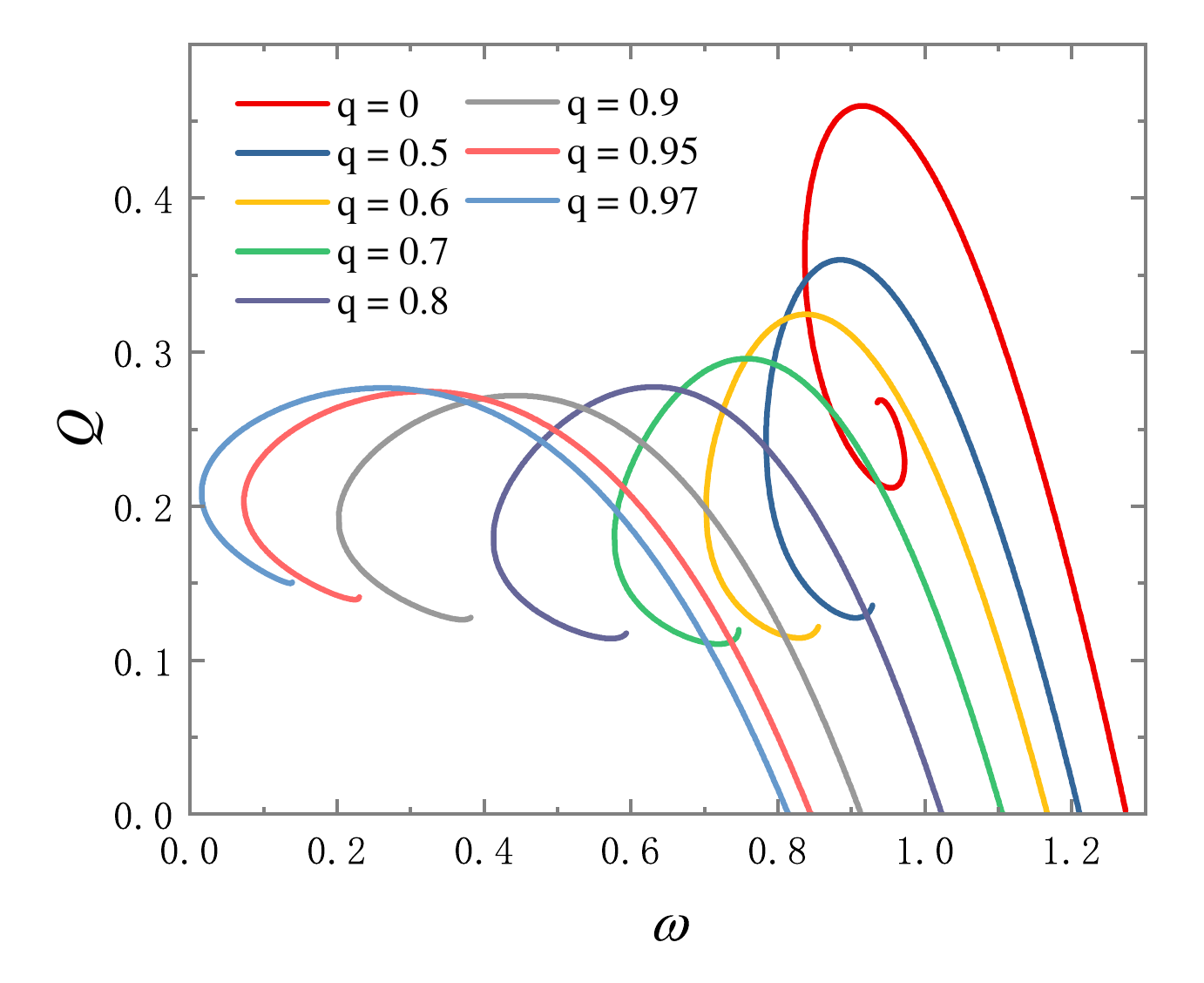}}
            \subfigure{\includegraphics[width=0.45\textwidth]{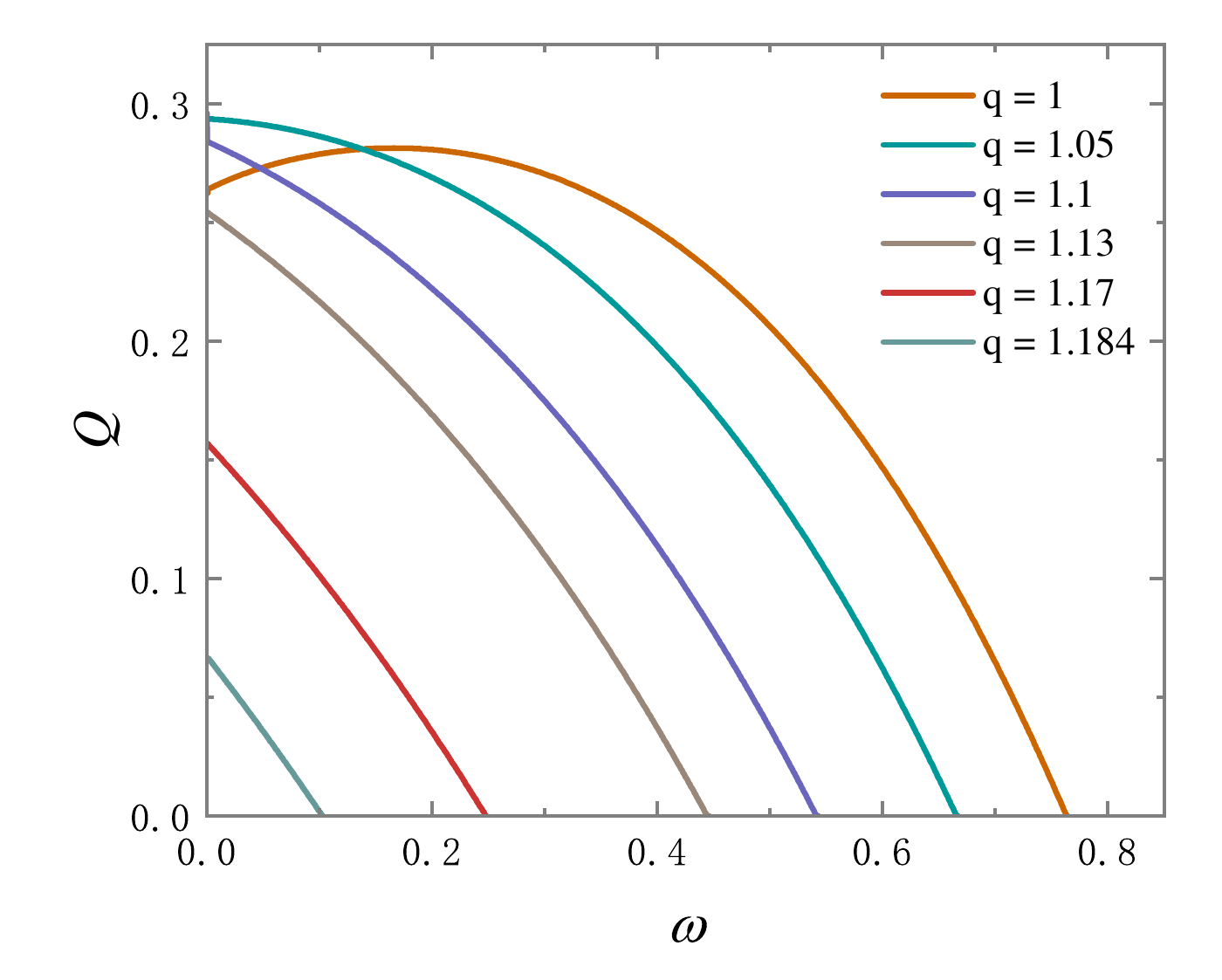}}
            \caption{The ADM mass $M$ and Noether charge $Q$ as functions of the frequency $\omega$ for different magnetic charges $q$. The cosmological constant is fixed at $\Lambda=-0.1$.}
            \label{p1}
          \end{figure}
By solving equations (\ref{eom15}-\ref{eom18}), we obtain solutions that satisfy the boundary conditions. Fig. \ref{p1} shows the variation of the ADM mass $M$ and Noether charge $Q$ of BDSs with different magnetic charges $q$ as a function of frequency $\omega$ in AdS spacetime ($\Lambda=-0.1$). The left column of the panels shows cases with smaller values of $q$. We can observe that the curves for $M$ and $Q$ exhibit multiple branches. The case with $q=0$ corresponds to a spherically symmetric Dirac stars. The frequency of the diluted solutions where $M$ and $Q$ approach zero is approximately $\omega=1.274$, and this value is not exactly one due to the influence of the cosmological constant \cite{Zhang:2024nju}. The minimum value of $M$ gradually becomes non-zero as $q$ increases, while the minimum value of $Q$ remains zero. The magnitude of $Q$ can represent the number of particles, so $Q=0$ indicates that the number of fermions is zero. In this case, the Bardeen term contributes to the $M$ of the BDSs. The entire curve shifts to lower frequency values as $q$ increases. It can be observed that the curve for $q=0.97$ is already very close to $\omega=0$. The right column of the panels shows cases with larger values of $q$. The curves gradually shorten as $q$ increases. The minimum frequencies for $M$ and $Q$ approach zero, and the second branch no longer appears. Additionally, $M_{max}$ of the BDSs also increases as $q$ grows. Larger magnetic charges correspond to larger ADM masses for the BDSs. However, the behavior of $Q_{max}$ is different. It does not vary monotonically with $q$.

          \begin{figure}
            \centering
            \subfigure{\includegraphics[width=0.38\textwidth]{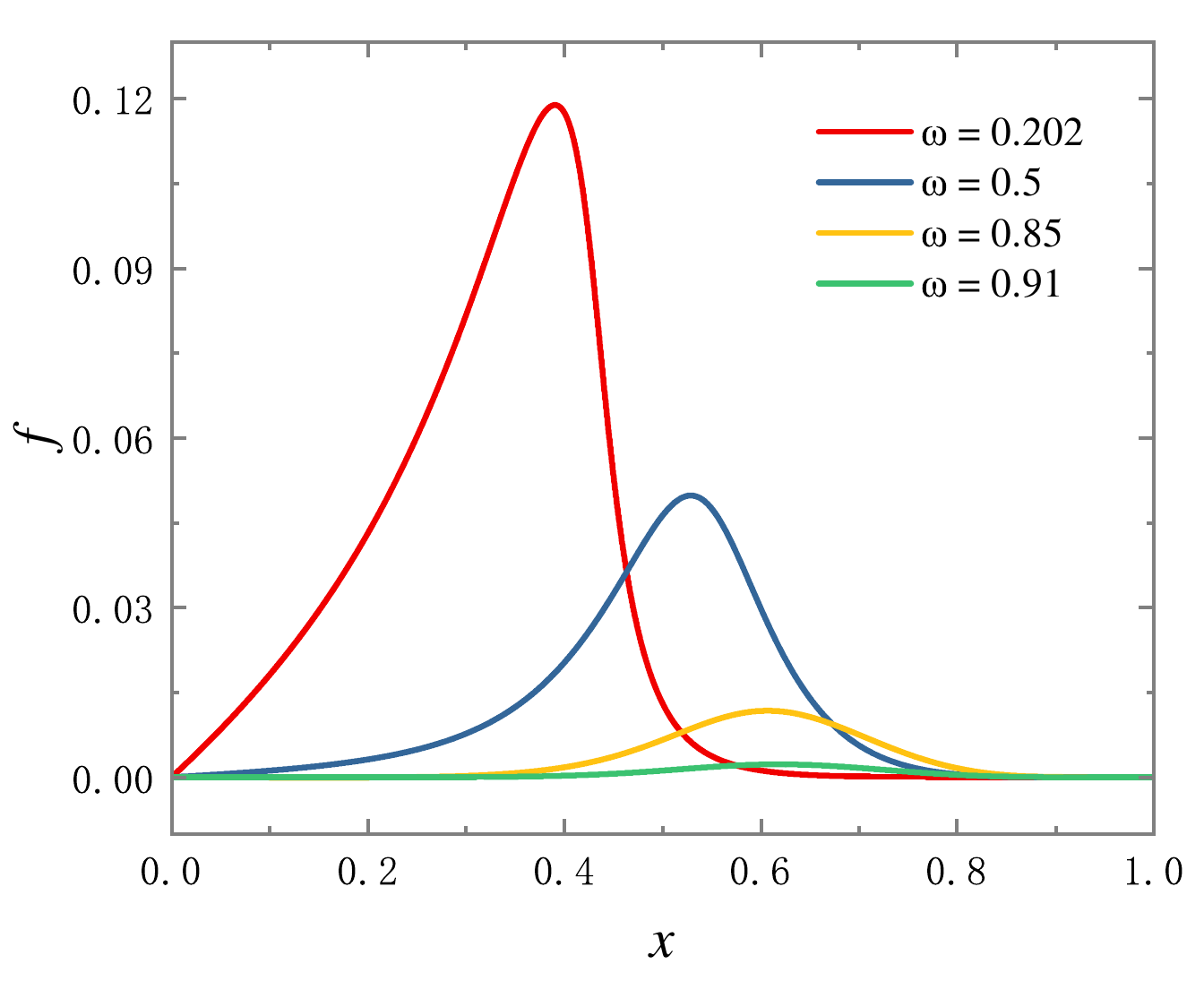}}
            \subfigure{\includegraphics[width=0.38\textwidth]{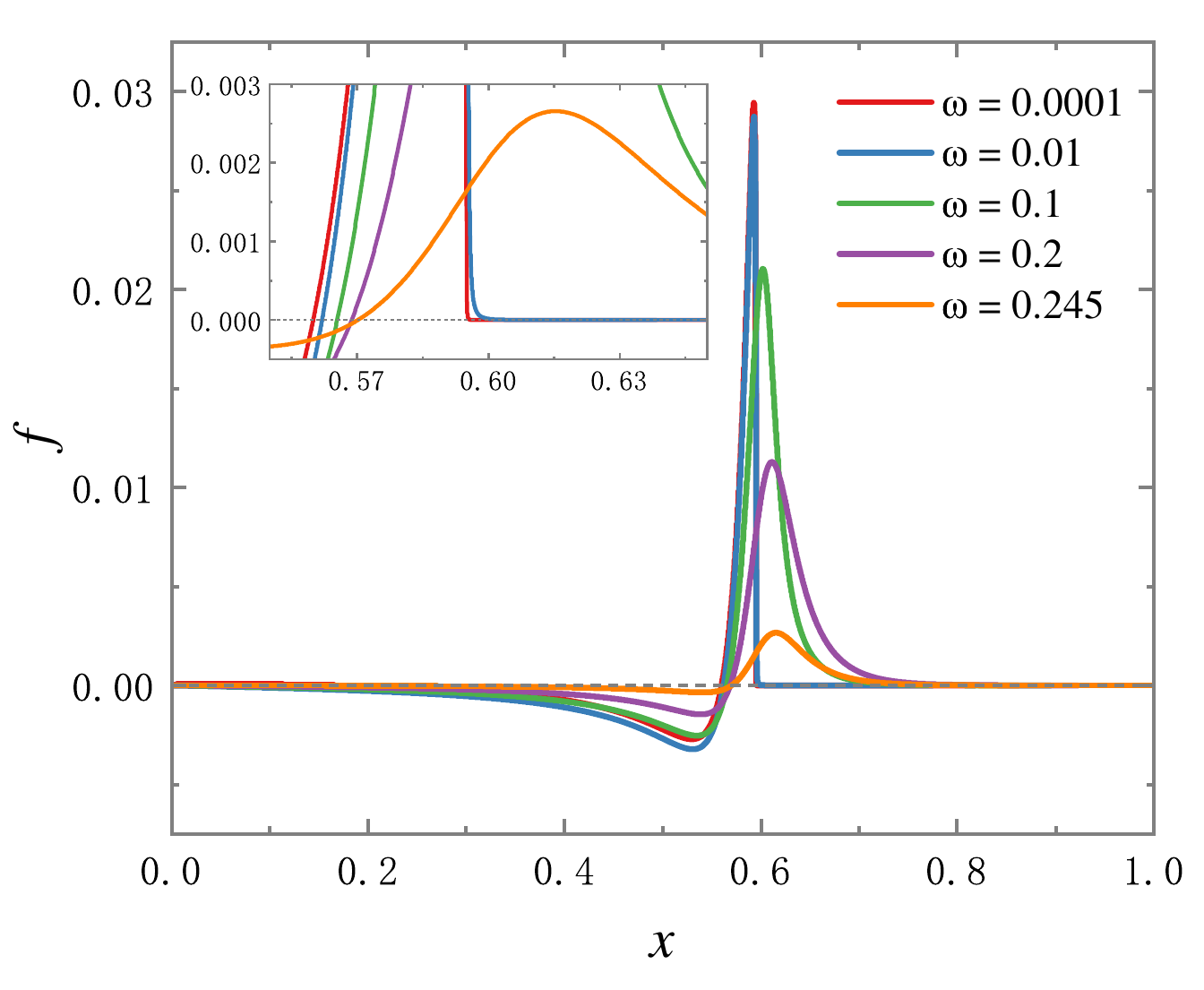}}
            \subfigure{\includegraphics[width=0.38\textwidth]{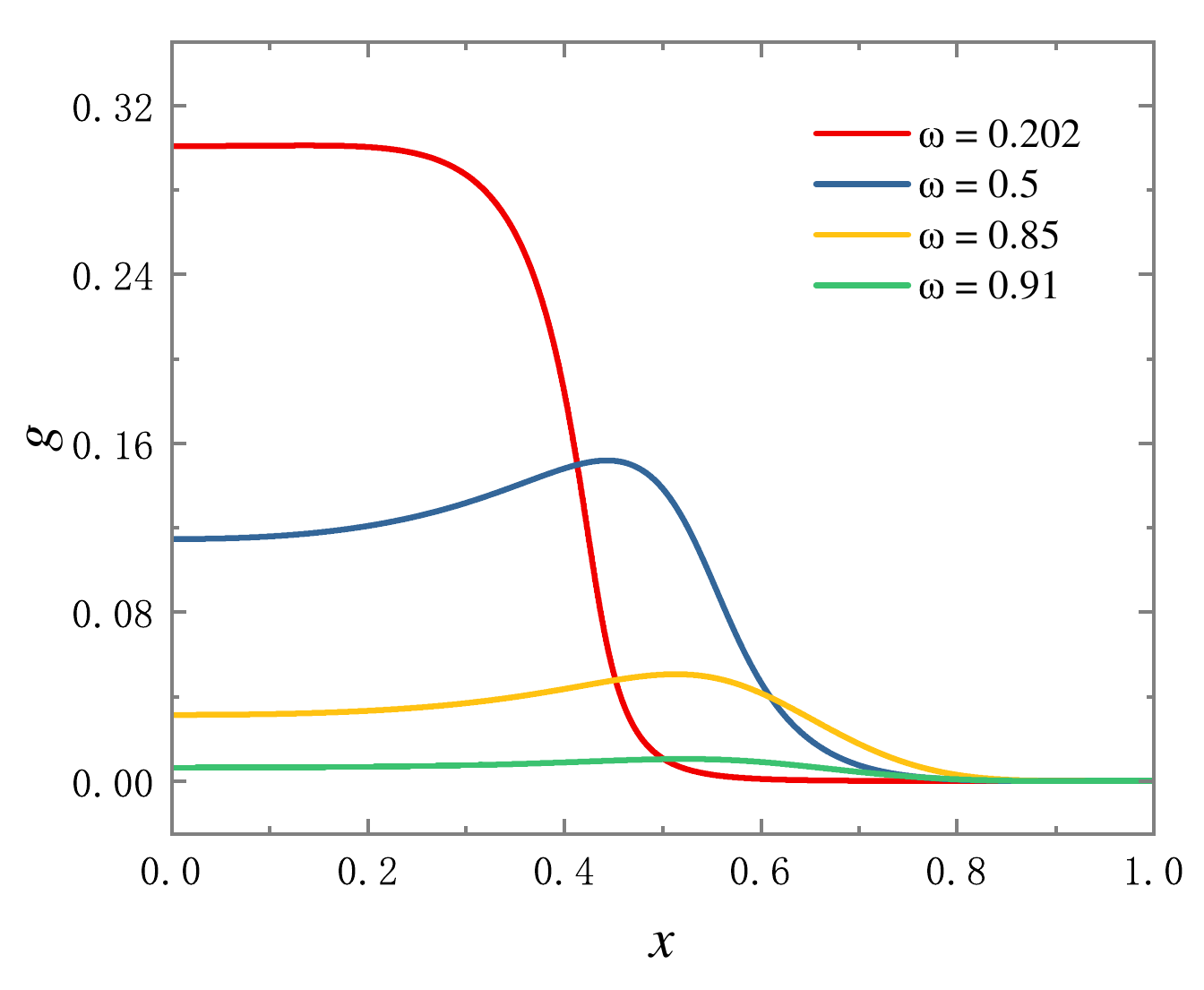}}
            \subfigure{\includegraphics[width=0.38\textwidth]{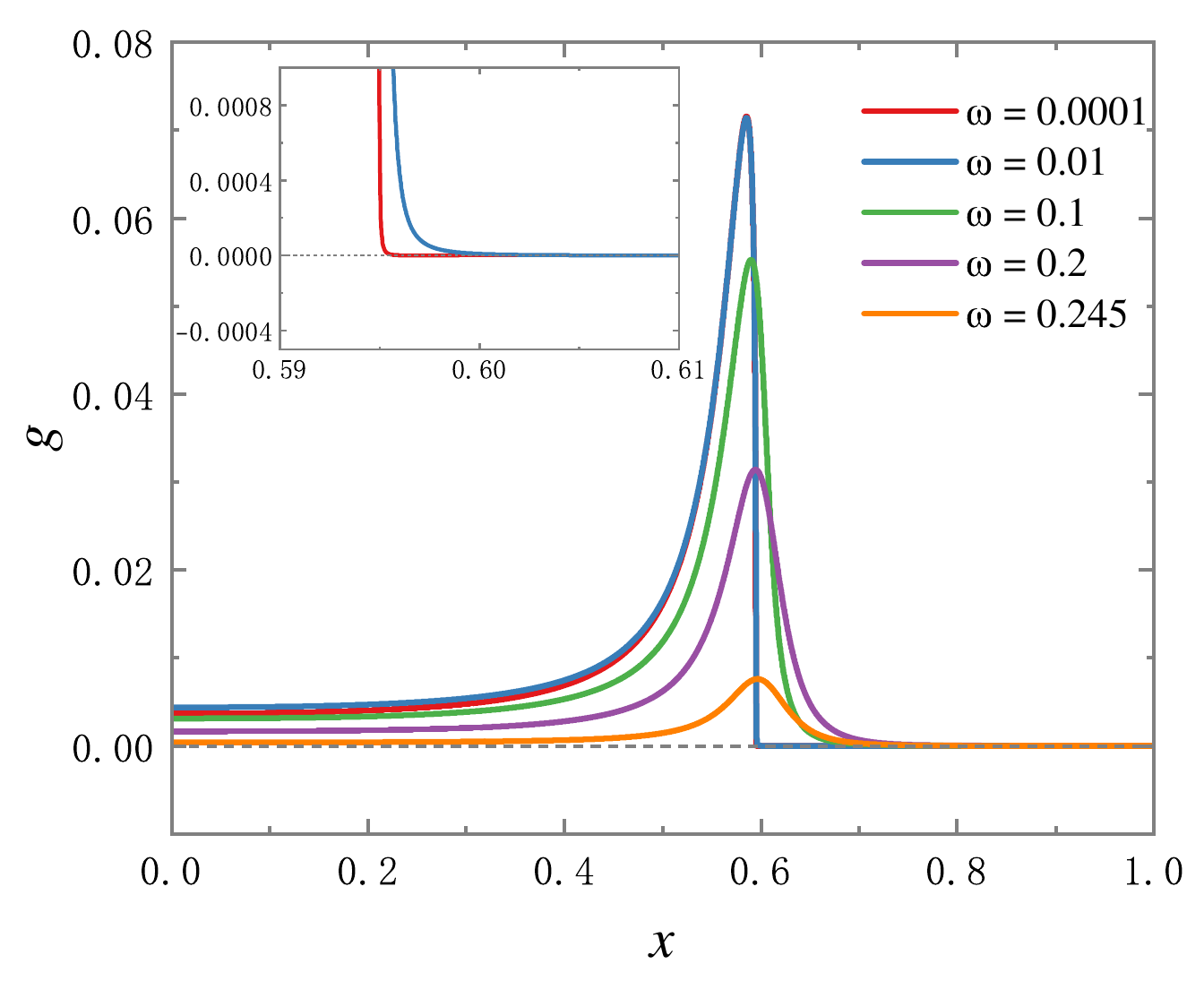}}
            \subfigure{\includegraphics[width=0.38\textwidth]{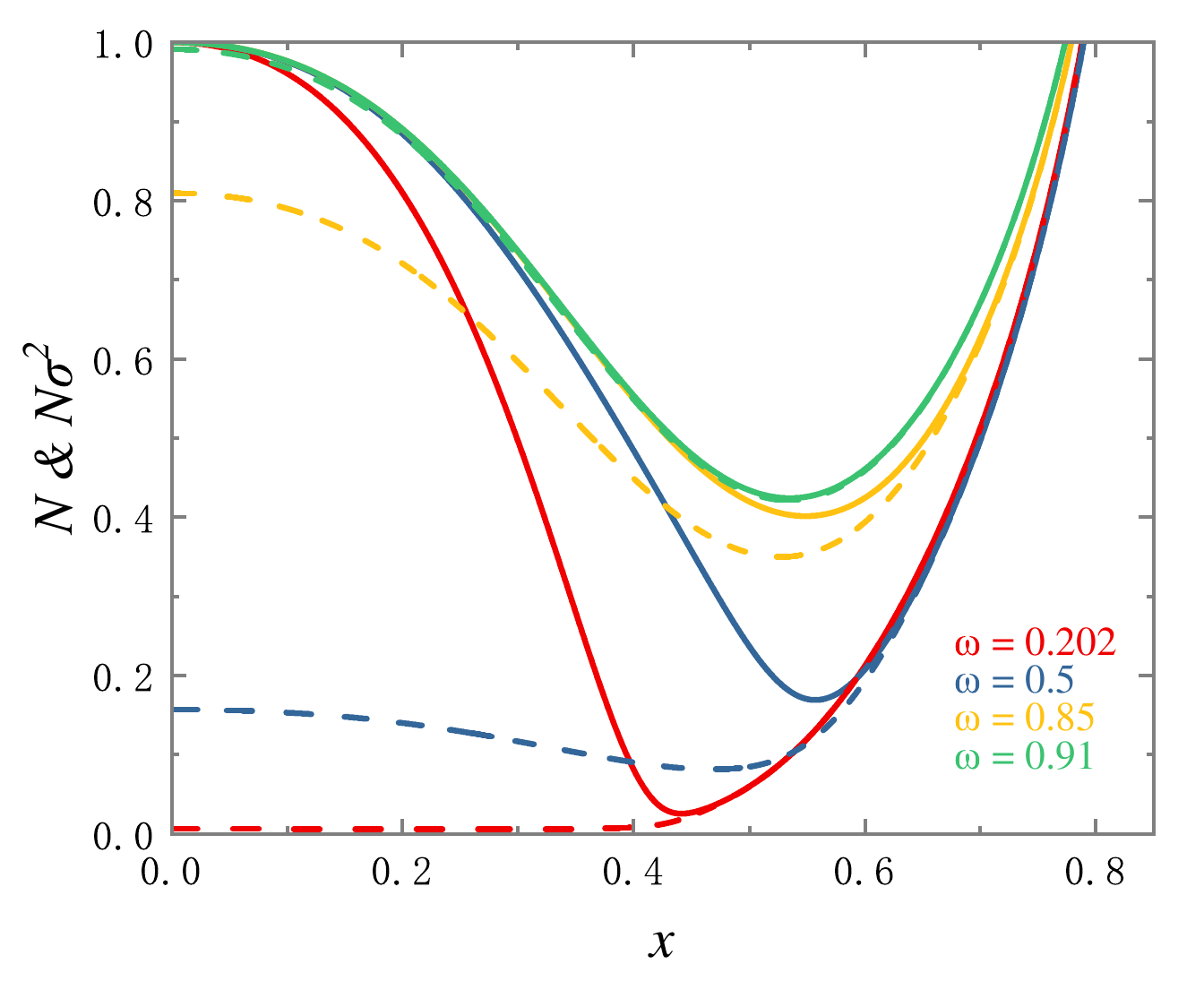}}
            \subfigure{\includegraphics[width=0.38\textwidth]{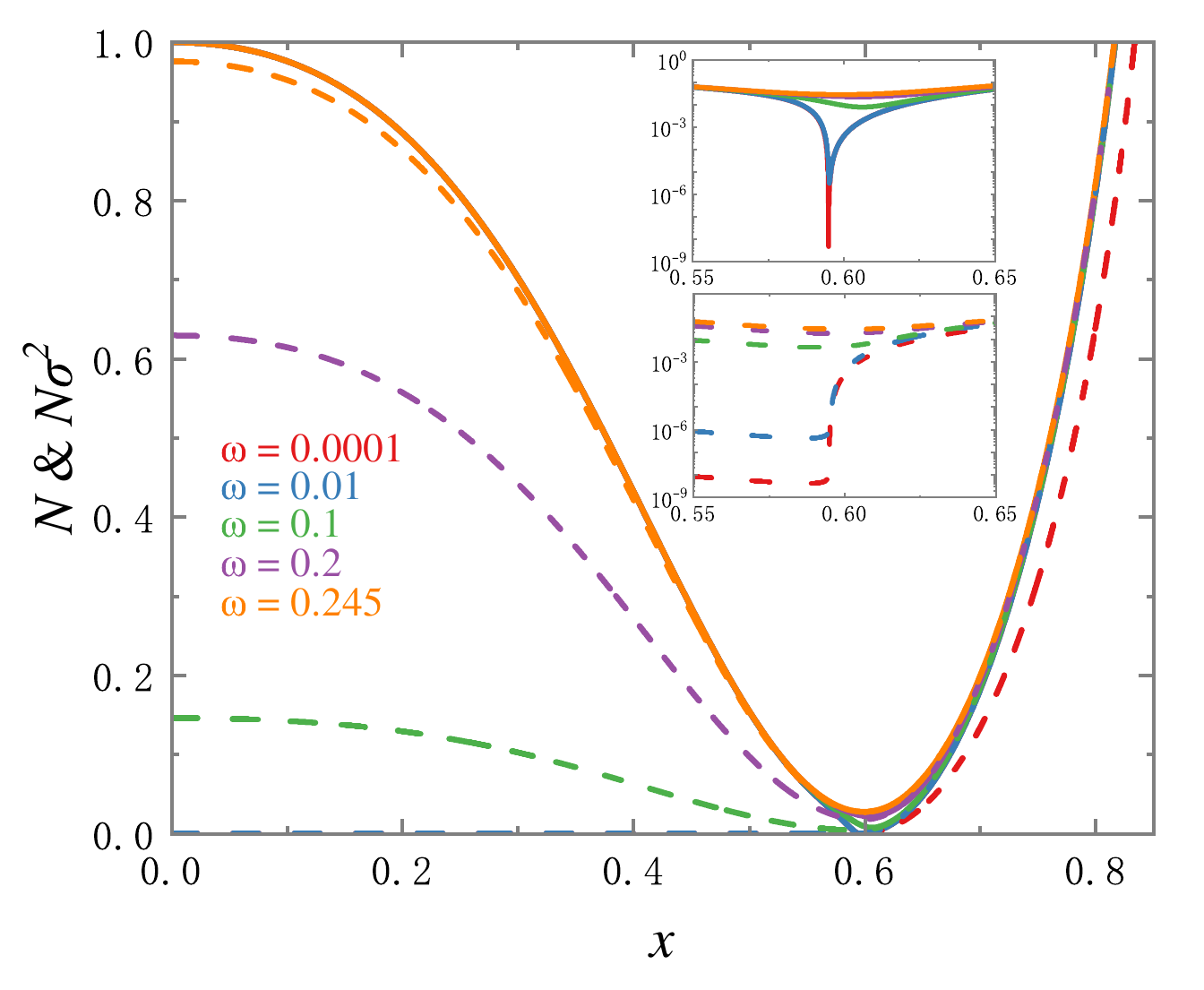}}
            \subfigure{\includegraphics[width=0.38\textwidth]{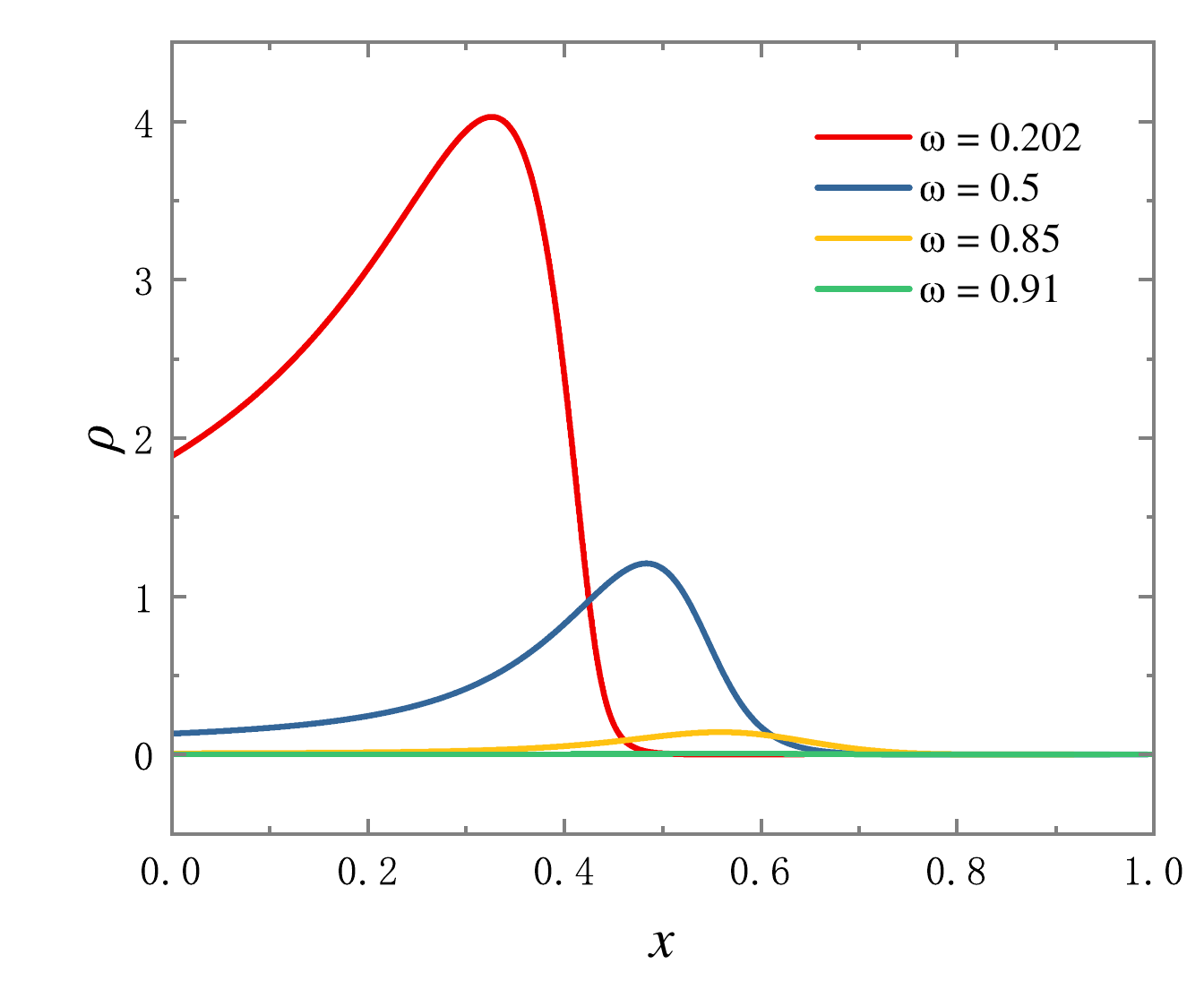}}
            \subfigure{\includegraphics[width=0.38\textwidth]{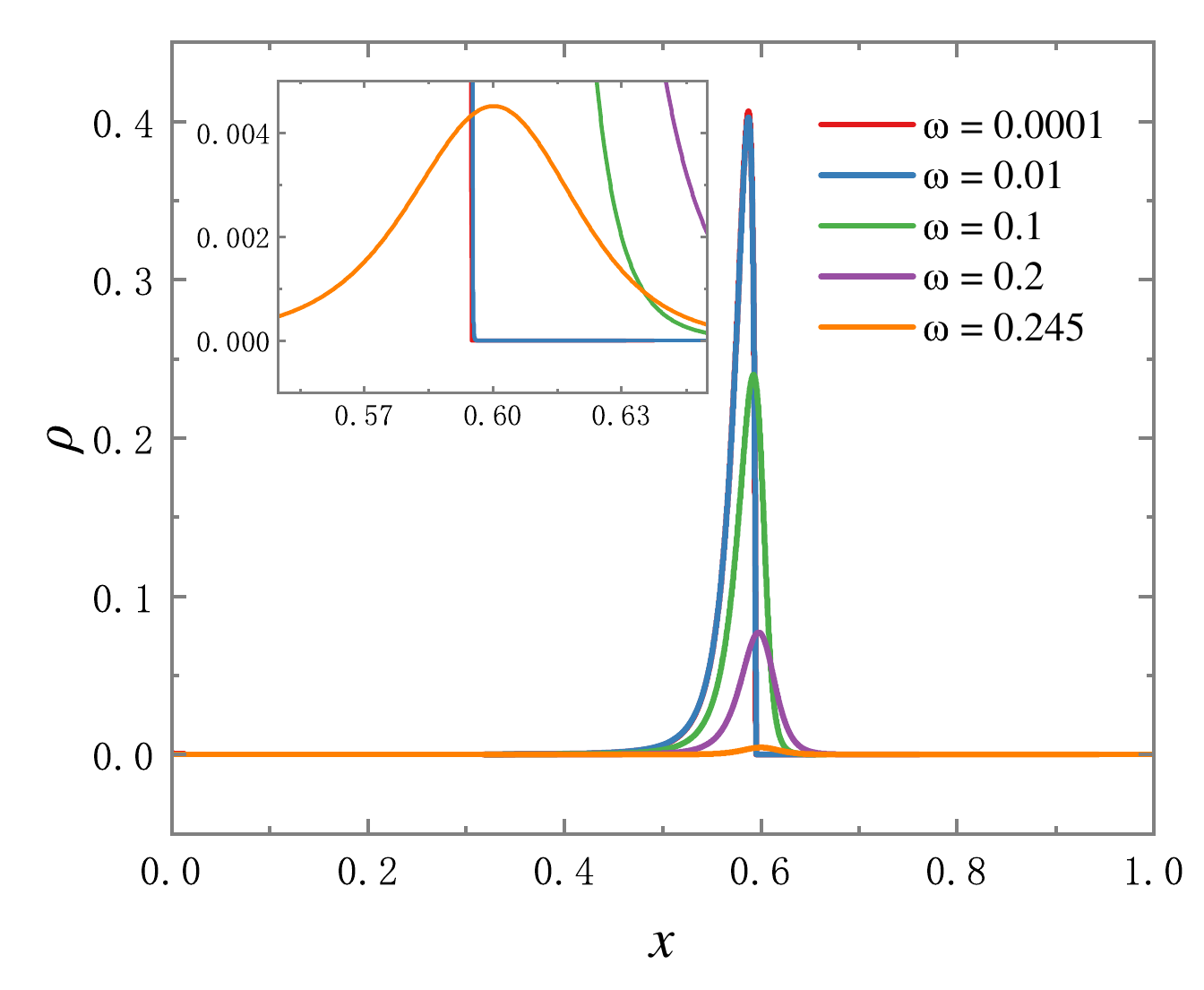}}
            \caption{The Dirac field functions $f$ and $g$, metric functions  $N$ and $N\sigma^{2}$, and energy density $\rho$ as functions of $x$ for multiple frequencies $\omega$. Solid lines represent $N$, and dashed lines represent $N\sigma^ {2}$. The left column corresponds to $q=0.9$, and the right column corresponds to $q=1.17$. The cosmological constant is fixed at $\Lambda=-0.1$.}
            \label{p2}
        \end{figure}
\begin{figure}
        \centering
        \begin{minipage}{0.45\textwidth}
          \centering
          \includegraphics[width=\textwidth]{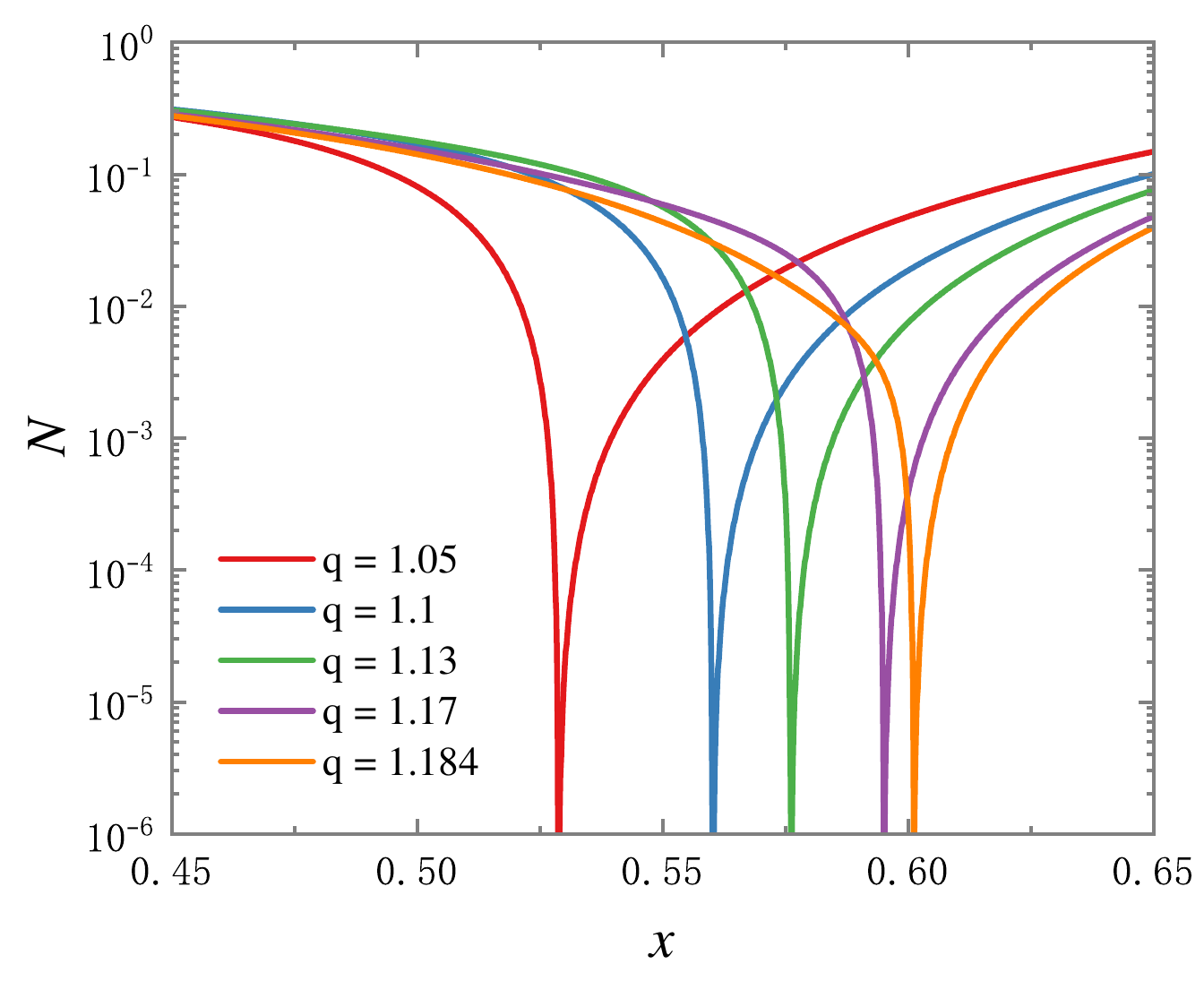}
          \caption{The metric function $N$ as a function of $x$ for different magnetic charges $q$. The cosmological constant and the frequency is fixed at $\Lambda=-0.1$, $\omega=0.0002$.}
          \label{p3}
        \end{minipage}
        \begin{minipage}{0.5\textwidth}
          \centering
          \begin{tabular}{>{\centering\arraybackslash}m{1cm}||>{\centering\arraybackslash}m{1cm}|>{\centering\arraybackslash}m{1cm}|>{\centering\arraybackslash}m{1cm}|>{\centering\arraybackslash}m{1cm}|>{\centering\arraybackslash}m{1cm}}
            \hline
            $q$ & 1.05 & 1.1 & 1.13 & 1.17 & 1.184 \\[10pt]
            \hline
            $x_{cH}$ & 0.529 & 0.560 & 0.576 & 0.595 & 0.601 \\[10pt]
            \hline
            $r_{cH}$ & 1.123 & 1.274 & 1.360 & 1.470 & 1.508 \\[10pt]
            \hline
          \end{tabular}
          \captionof{table}{The critical horizon positions $x_{cH}$ and $r_{cH}$ for different values of $q$.}
          \label{tab1}
        \end{minipage}
        \end{figure}

In Fig. \ref{p2}, we show the field functions $f$ and $g$, the metric functions $N$ and $N\sigma^{2}$ and the energy density $\rho$ for $q=0.9$ (left column) and $q=1.17$ (right column). In the case of $q=0.9$, $\omega=0.91$ (green line) corresponds to the rightmost end of the curve. It can be observed that the fields $f,g$ and the energy density $\rho$ are close to zero throughout the entire space. This also indicates that the contribution of the Dirac field to the mass is nearly zero at this point. The maximum values of the fields $f_{max},g_{max}$ and the maximum value of the energy density $\rho_{max}$ gradually increase as $\omega$ decreases. Meanwhile, the minimum values of the metric functions $N_{min},N\sigma^{2}_{min}$ gradually decrease. When $q=0.202$, which corresponds to the leftmost end of the curve, or the end of the first branch, the value of $N_{min}$ is approximately $0.006$ and $N\sigma^{2}_{min}$ is approximately $0.025$. When $q=1.17$, the trends for the field functions, metric functions and energy density are similar as $\omega$ decreases. At $\omega=0.0001$, the field functions $f,g$ and the energy density $\rho$ become very steep, forming a barrier similar to a spherical shell. Outside this barrier, there is almost no matter present. This solution is referred to as the Frozen Bardeen-boson stars (FBDSs). The inset shows details around $x=0.6$. At this point, $N_{min},N\sigma^{2}_{min}$ are very small, on the order of $10^{-9}$ , and are very close to zero. This property is similar to the event horizon of a black hole, which confines matter within $x_{cH}$. This location $x_{cH}$ is defined as the “critical horizon".
In Fig. \ref{p3}, we study the metric function $N$ of BDSs with different $q$ in AdS spacetime. We find that the critical horizon position $x_{cH}$ also gradually increases as $q$ increases. Table \ref{tab1} presents the specific values of $x_{cH}$ and $r_{cH}$.

\begin{figure}
            \centering
            \subfigure{\includegraphics[width=0.45\textwidth]{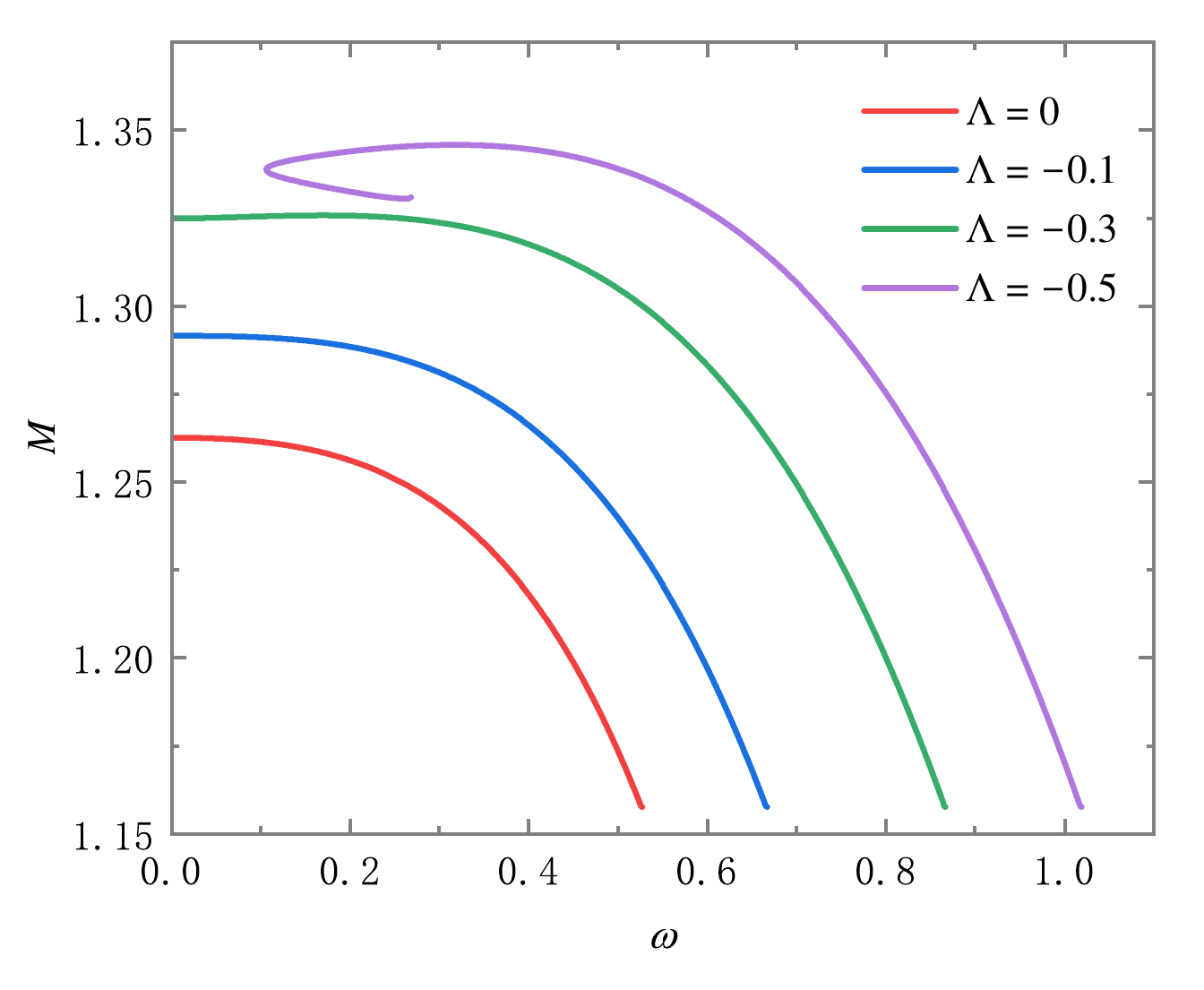}}
            \subfigure{\includegraphics[width=0.45\textwidth]{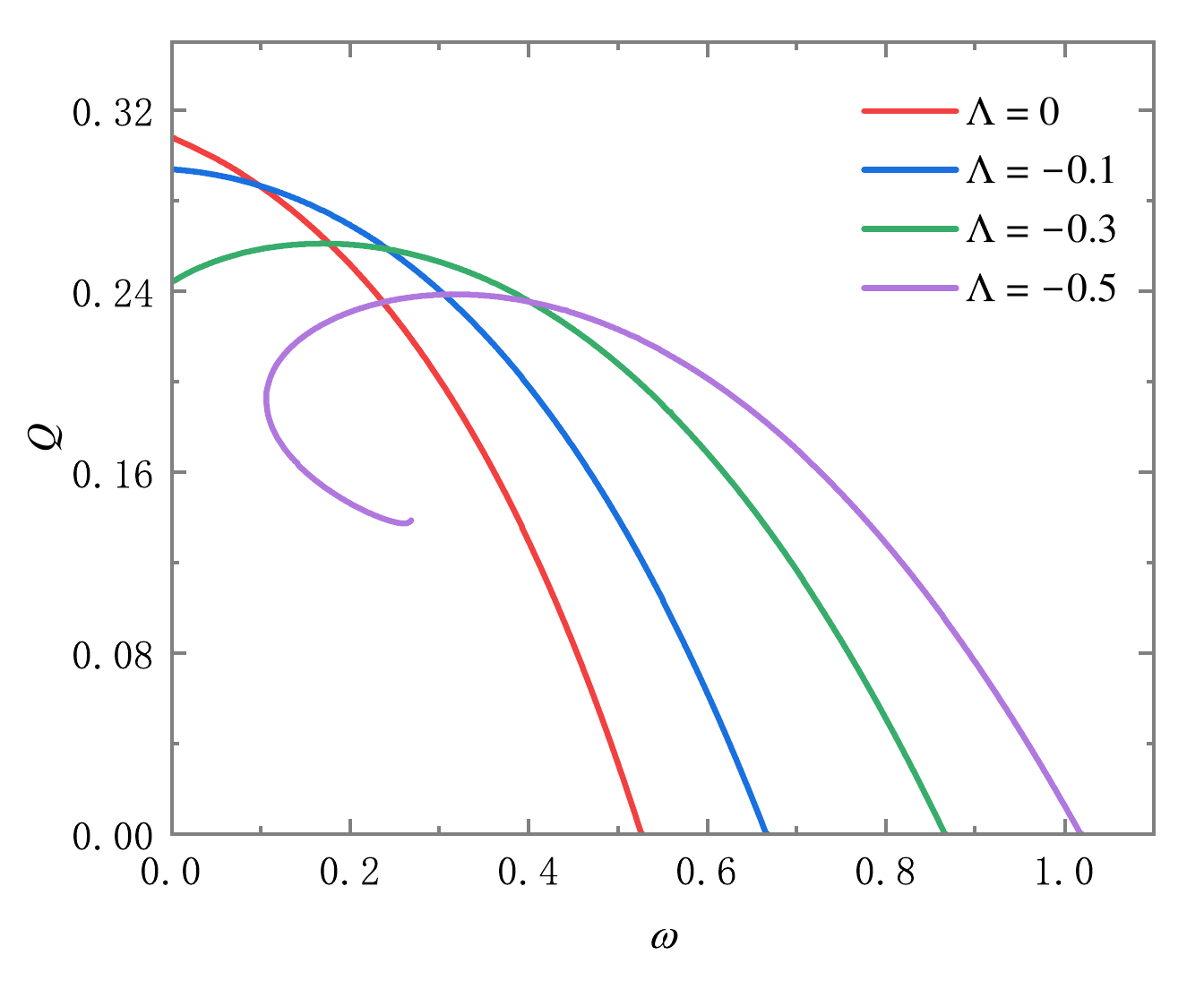}}
            \caption{The ADM mass $M$ (left panel) and Noether charge $Q$ (right panel) as functions of frequency $\omega$ for different cosmological constants $\Lambda$, with $q=1.05$.}
            \label{p4}
        \end{figure}

\begin{figure}
            \centering
            \subfigure{\includegraphics[width=0.31\textwidth]{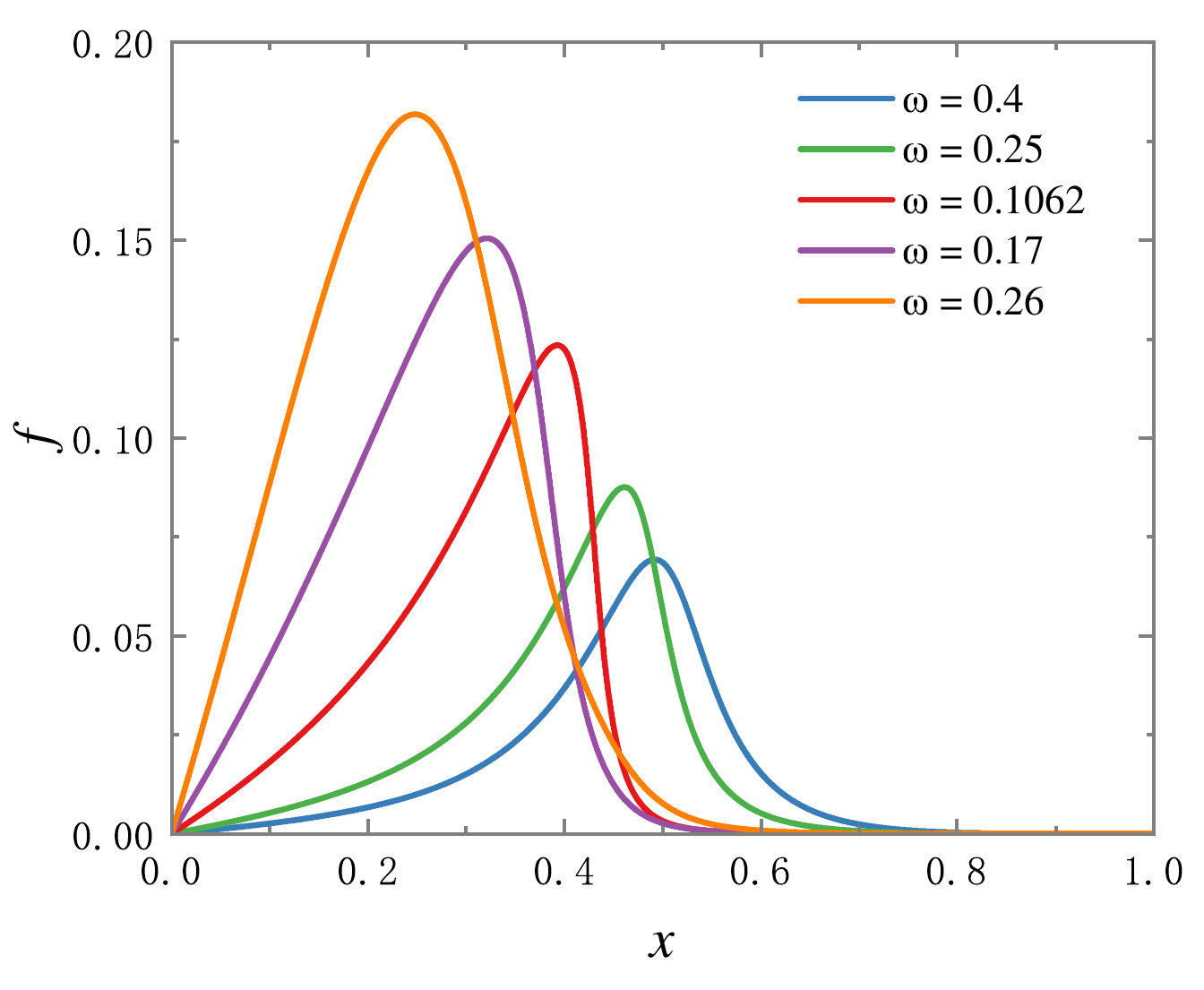}}
            \subfigure{\includegraphics[width=0.31\textwidth]{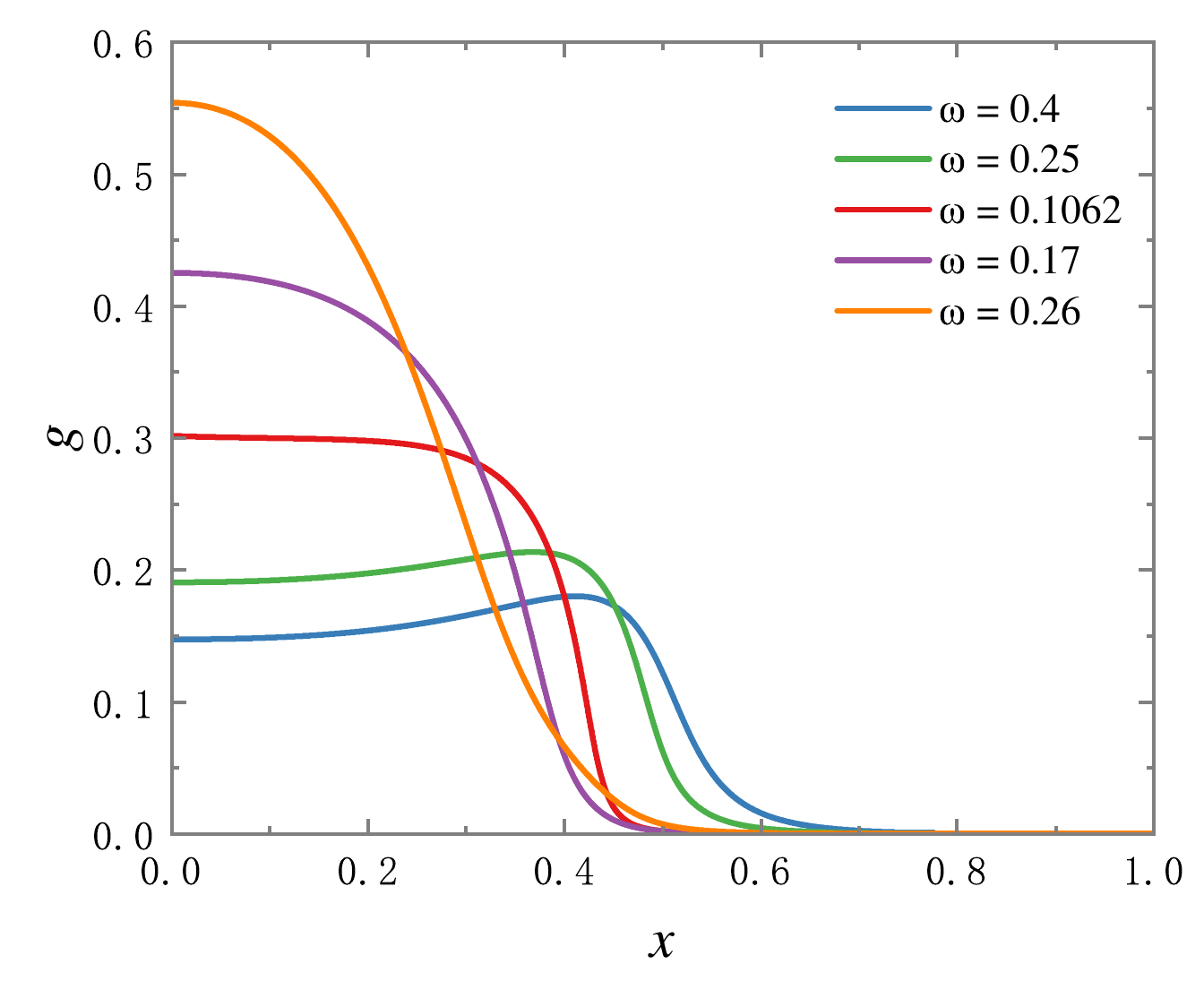}}
            \subfigure{\includegraphics[width=0.31\textwidth]{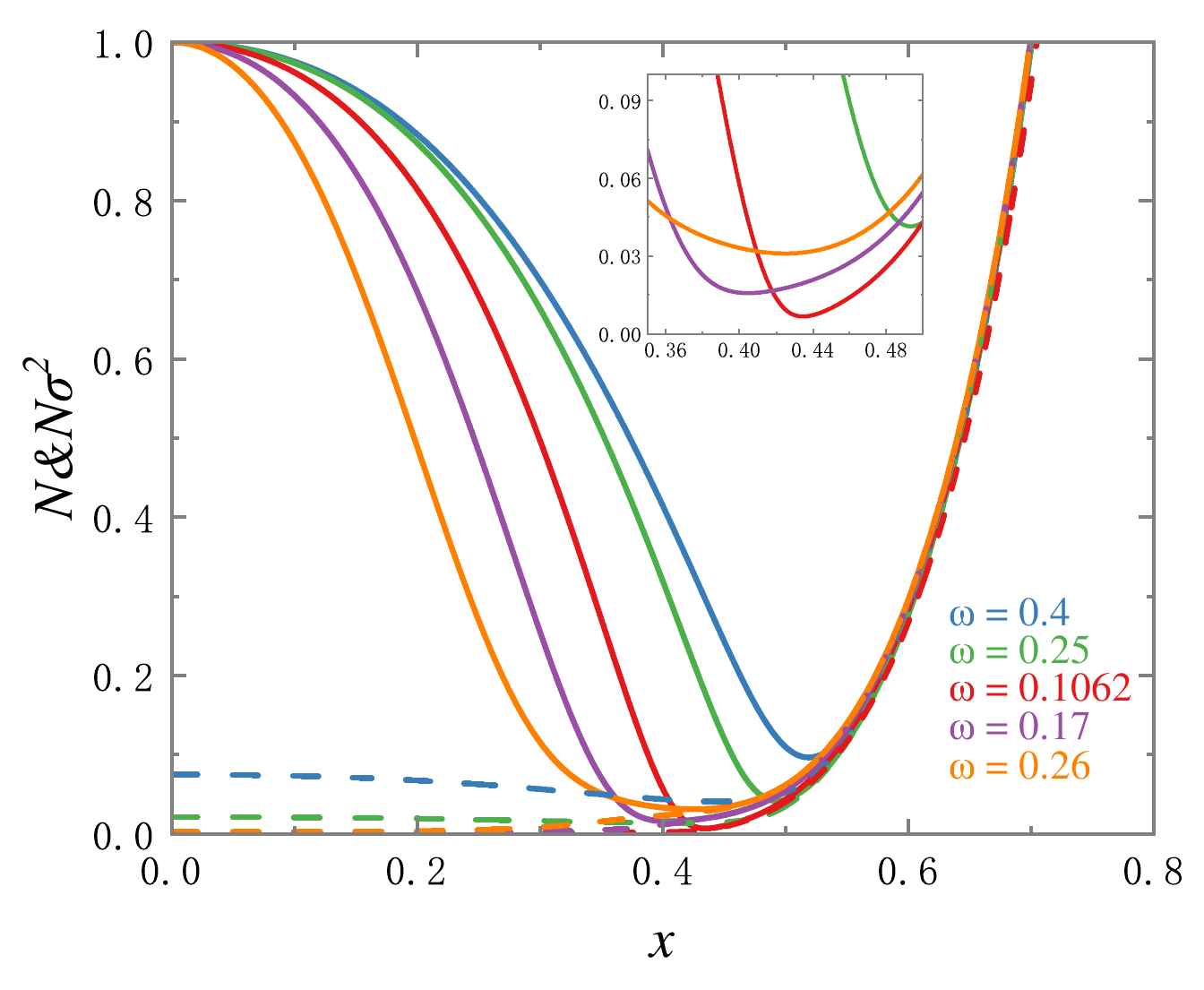}}
            \caption{The field functions $f$ and $g$, and the metric function $N$ and $N\sigma^{2}$ as functions of $x$ with different frequencies $\omega$.  The cosmological constant and the magnetic charge is fixed at $\Lambda=-0.5$, $q=1.05$.}
            \label{p5}
        \end{figure}

In the previous discussions, we fixed the cosmological constant. The properties of BDSs in AdS spacetime are similar to those in asymptotically Minkowski spacetime, though there are numerical differences. Next, we will set different values for the cosmological constant to study its impact on BDSs. Fig. \ref{p4} shows the ADM mass $M$ and Noether charge $Q$ of BDSs with $q=1.05$ under different cosmological constants. $M_{max}$ gradually increases as the cosmological constant decreases. $M_{min}$ does not change with $\Lambda$, but the corresponding frequency increases, which is still due to the influence of the cosmological constant. In contrast, the behavior of the Noether charge $Q$ is different. $Q_{max}$ decreases as the cosmological constant decreases while $Q_{min}$ remains unchanged. This indicates that, due to the influence of the cosmological constant, the Dirac field's contribution to the mass $M$ gradually decreases. However, the influence of the cosmological constant becomes less significant as $M$ becomes smaller.  At the point $M_{min}$, the cosmological constant does not affect the proportion of the Dirac field's contribution to the total mass. We also find that decreasing the cosmological constant causes the curves of $M$ and $Q$ exhibit new branches, and the minimum frequency of the curves increases. In Fig. \ref{p5}, we present the field functions and metric functions of BDSs with $\Lambda=-0.5,q=1.05$, where no frozen solutions are found. In the discussion of Fig. \ref{p2}, we observed that BDSs with $\Lambda=-0.1$ exhibited freezing, meaning there was a critical horizon $x_{cH}$ that confined the matter within a shell. However, as the cosmological constant decreases, this freezing effect gradually disappears. This indicates that frozen solutions are more likely to occur for BDSs with values of $q$ closer to the limit, and a smaller $\Lambda$ requires a correspondingly larger $q$.

        \begin{figure}[b]
            \centering
            \subfigure{\includegraphics[width=0.45\textwidth]{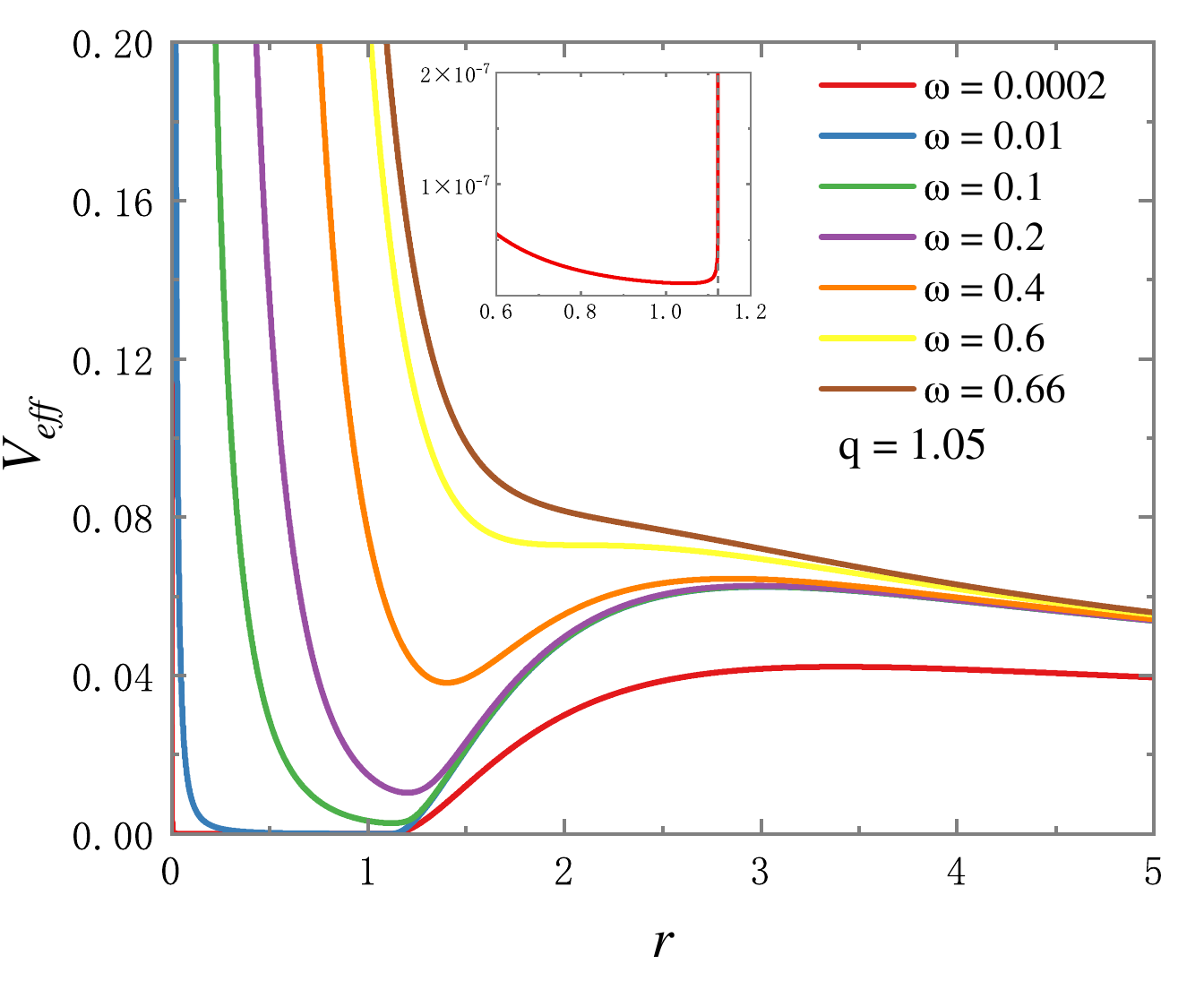}}
            \subfigure{\includegraphics[width=0.45\textwidth]{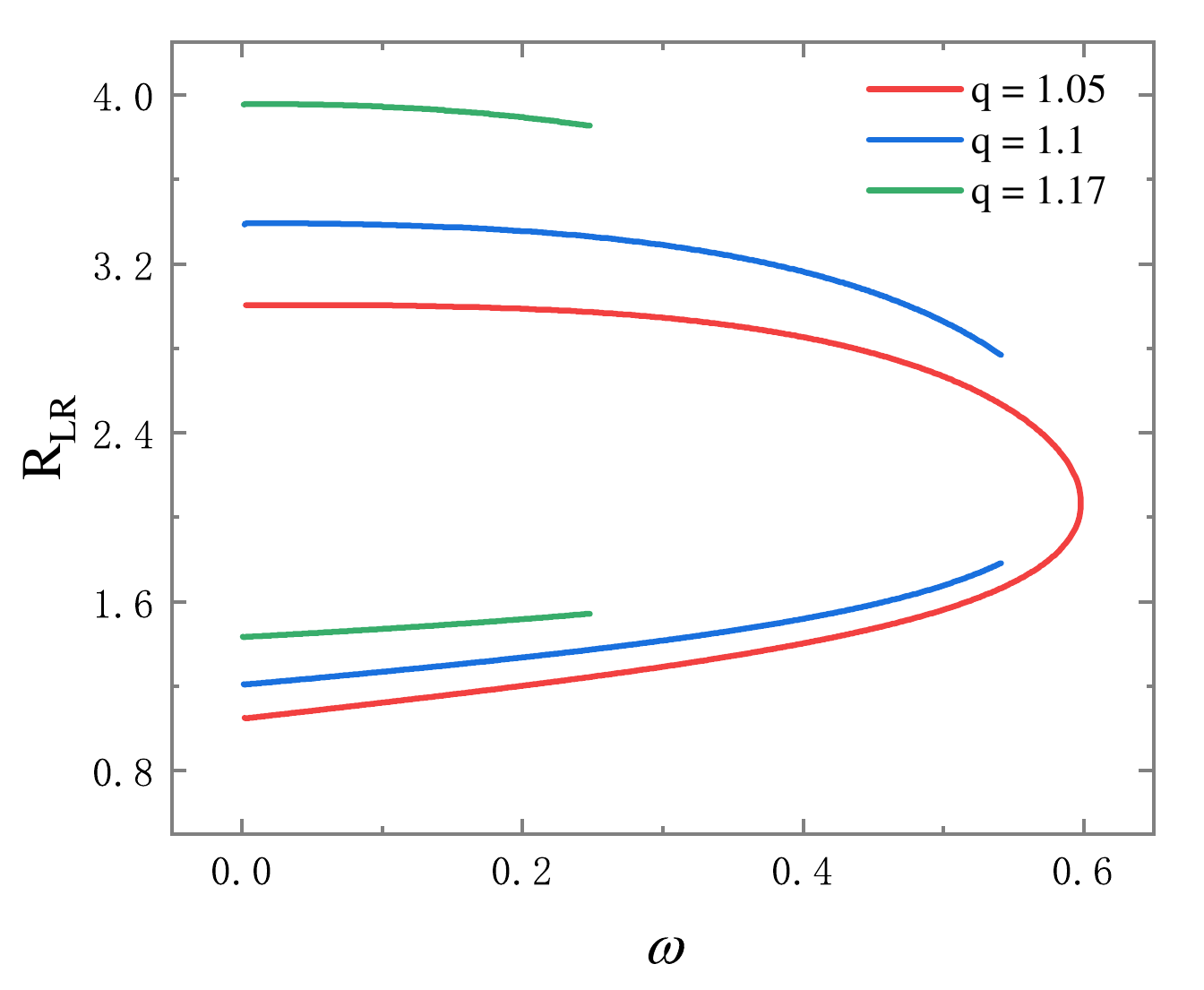}}
            \caption{Left panel: The effective potential $V_{eff}$ as a function of the radial coordinate $r$ for different frequencies $\omega$, with $q=1.05$. Right panel: The location of the light rings for BDSs $R_{LR}$ with different magnetic charges $q$ as a function of frequency $\omega$.  The cosmological constant is fixed at $\Lambda=-0.1$.}
            \label{p6}
        \end{figure}
Next, we will analyze the light rings of BDSs. In the previous Sec. \ref{c} , we derived the effective potential (\ref{eom23}). By numerically computing the specific solutions, we can determine the location of the light ring. Fig. \ref{p6} illustrates the variation of the effective potential $V_{eff}$ with the radial coordinate $r$ for BDSs with $\Lambda=-0.1,q=1.05$, as well as the dependence of the light ring position $R_{LR}$ on the frequency $\omega$ for BDSs with different values of $q$. In the left panel, we compute the effective potential curves for different frequencies, including $\omega=0.0002$ for FBDSs. We can observe that the effective potential develops two extrema as the frequency decreases. These two extrema correspond to two distinct light rings. If the second derivative of the effective potential is positive, the light ring is stable.  Otherwise, it is unstable. This implies that a local minimum of the extremum corresponds to a stable light ring, whereas a local maximum indicates an unstable light ring. From the Fig. \ref{p6}, we can see that the inner light ring is stable, while the outer light ring is unstable. Any small perturbation could cause photons to either escape the light ring orbit or be captured by the gravitational field of the BDSs. Additionally, we have established that the critical horizon radius $r_{cH}=1.123$, and we can observe that the inner ring is within the critical horizon. For black holes, physical information within the event horizon cannot be directly observed. However, physical effects within the critical horizon of BDSs are observable, making the study of the inner light ring particularly significant. The dependence of the light ring position on frequency for $q=1.05$ (red line) is shown in the right panel. The upper part of the curve represents the outer light ring, while the lower part represents the inner light ring. We observe that as the frequency increases, the two light rings approach each other until they merge at a critical frequency. We also computed scenarios for other magnetic charges. When $q$ increases to $q=1.1$, the two light rings do not merge into one as the frequency increases, and the frequency range for the light rings becomes smaller. When $q$ increases further to $q=1.17$, the minimum distance between the two light rings also increases, and the frequency range becomes even smaller.

        \begin{figure}
            \centering
            \subfigure{\includegraphics[width=0.45\textwidth]{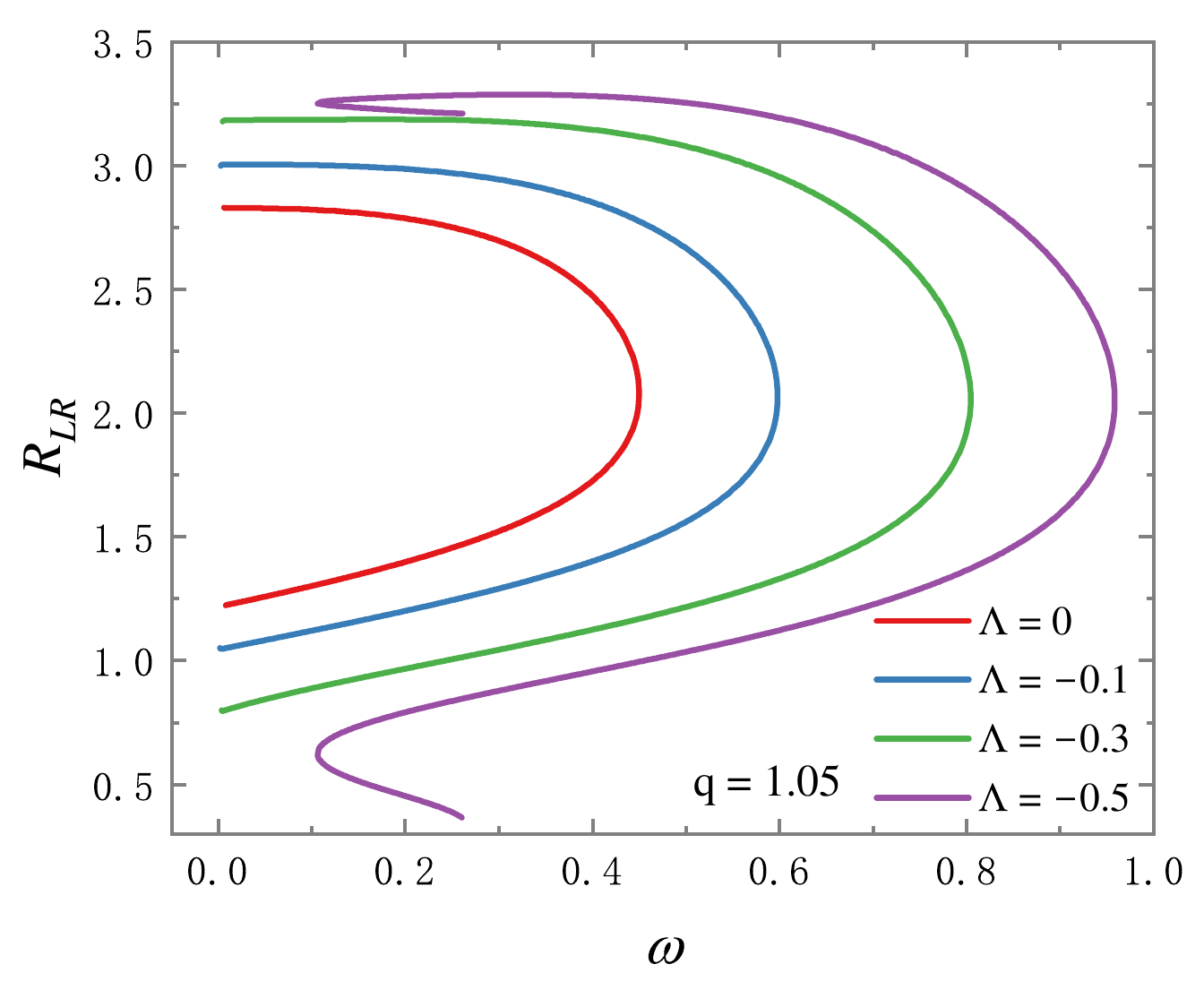}}
            \subfigure{\includegraphics[width=0.45\textwidth]{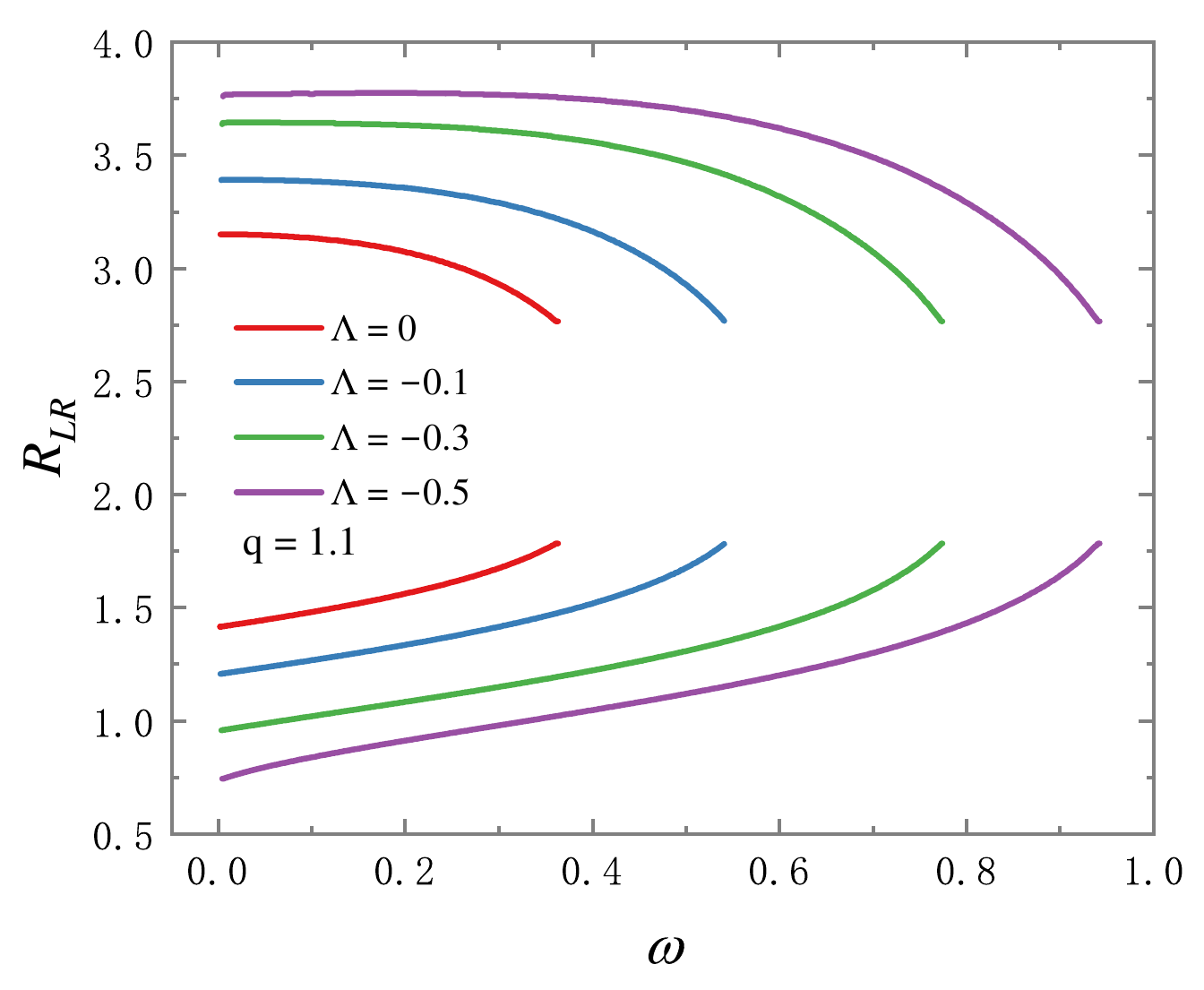}}
            \caption{The light ring positions of BDSs $R_{LR}$ with $q=1.05$ (left panel) and $q=1.1$ (right panel) under different cosmological constants $\Lambda$ as functions of frequency $\omega$.}
            \label{p7}
        \end{figure}
Similarly, we aim to study the effect of the cosmological constant on the light rings of BDSs. Fig. \ref{p7} shows the dependence of the light ring position $R_ {LR}$ on frequency for $q=1.05$ (left panel) and $q=1.1$ (right panel) under different cosmological constants $\Lambda$. For BDSs with $q=1.05$, the light rings merge into one at a critical frequency. From the left panel, we observe that the cosmological constant does not change this property but affects the frequency range of the light rings. For $\Lambda=-0.5$ (purple line), the second branch of the light ring appears, corresponding to the second branch of BDSs for this cosmological constant. This second branch is also observed in Fig. \ref{p4}. For BDSs with $q=1.1$, the two light rings do not intersect in asymptotically  Minkowski spacetime. As the cosmological constant decreases, this property of the BDSs light rings remains unchanged.

\begin{figure}
            \centering
            \subfigure{\includegraphics[width=0.32\textwidth]{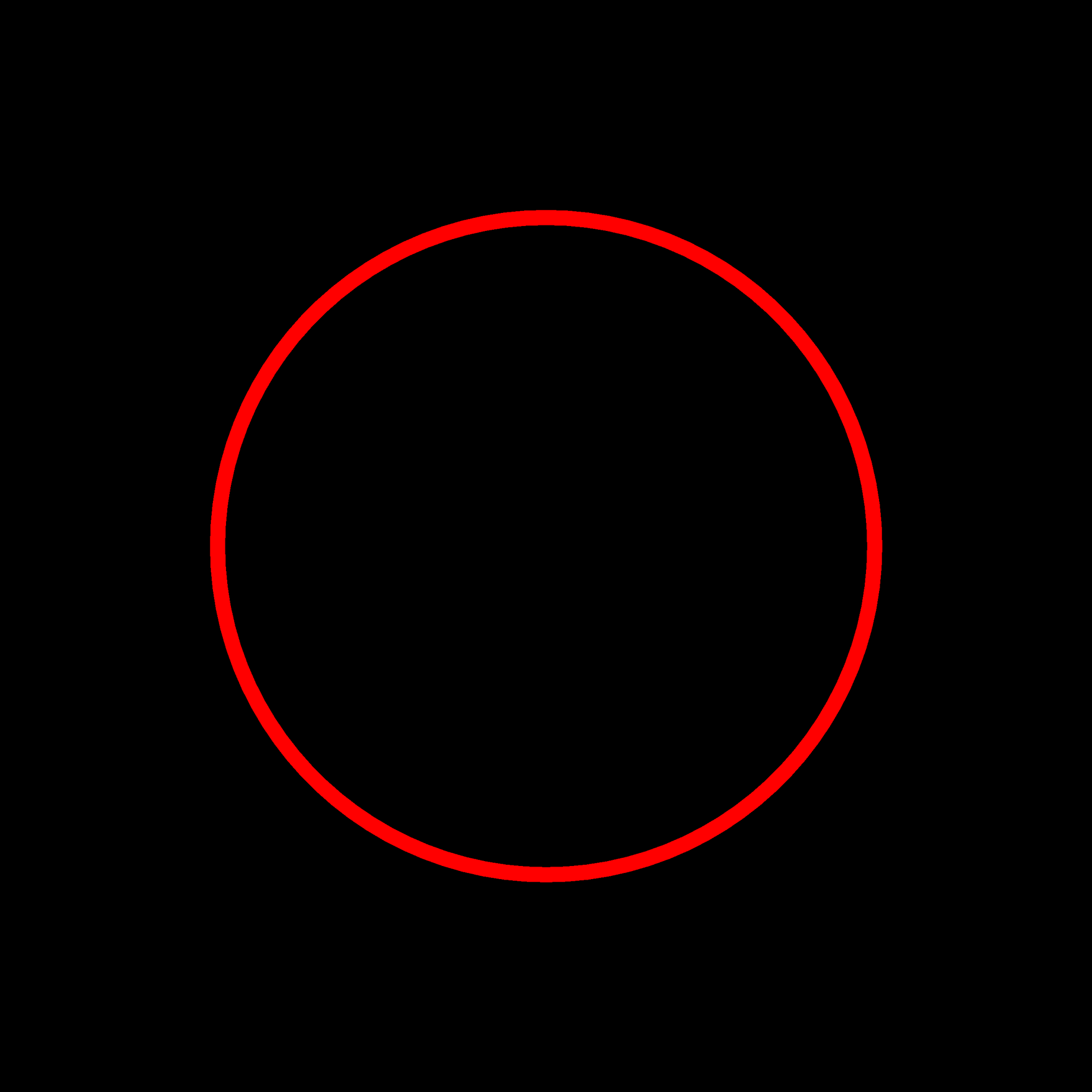}}
            \subfigure{\includegraphics[width=0.32\textwidth]{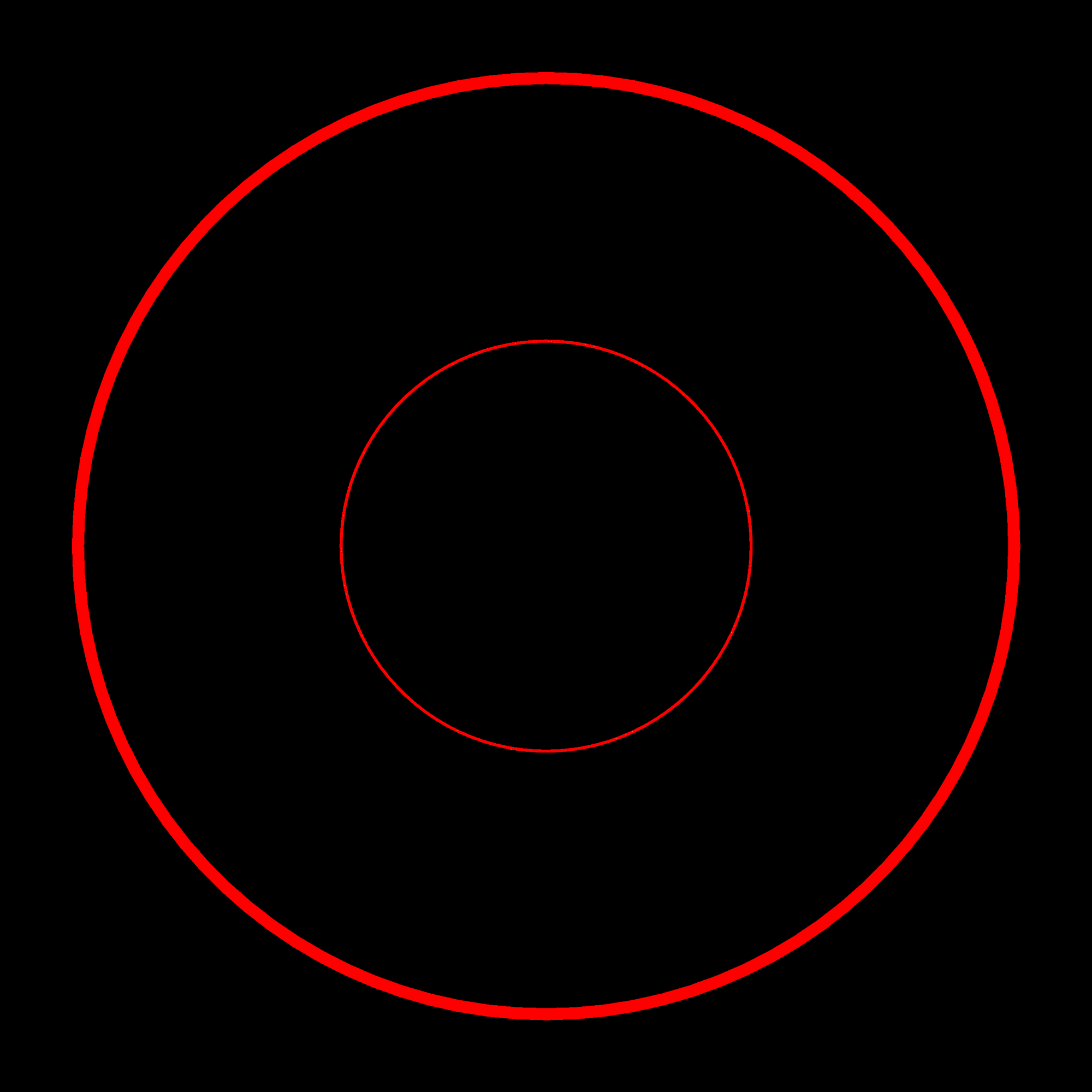}}
            \subfigure{\includegraphics[width=0.32\textwidth]{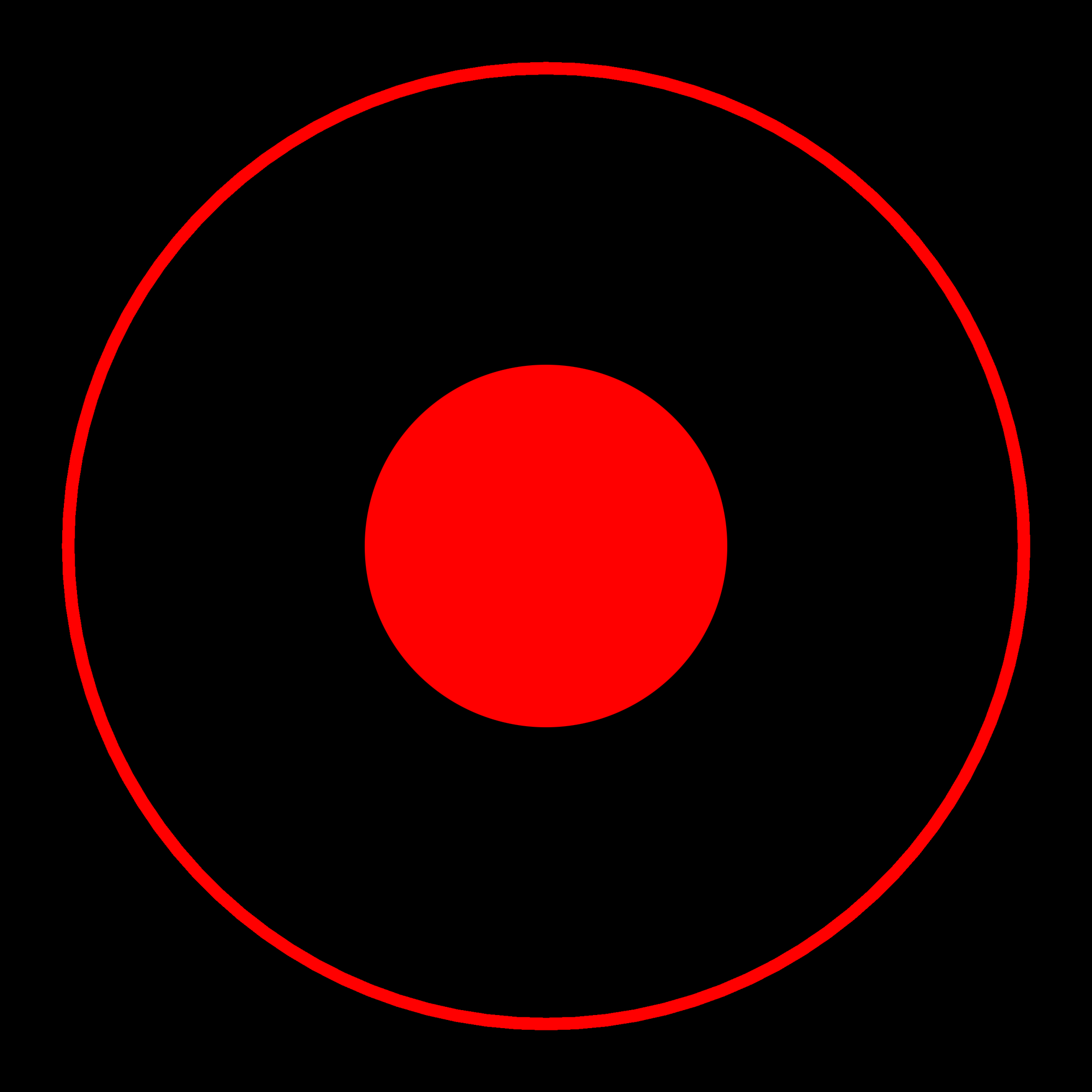}}
            \caption{The schematic diagrams of the light rings for frequencies $\omega=0.598,0.3,0.0002$, with the cosmological constant and magnetic charge fixed at $c=-0.1,q=1.05$. }
            \label{p8}
        \end{figure}

Fig. \ref{p8} illustrates the light rings for BDSs with $\Lambda=-0.1,q=1.05$. We have computed solutions for three different frequencies $\omega$. From left to right, the frequency corresponds to $\omega=0.598,0.3,0.0002$. For $\omega=0.598$ (left panel), the two light rings intersect, but only one light ring is visible in the plot. For $\omega=0.3$ (middle panel), both the inner and outer light rings are observable. When $\omega=0.0002$ (right panel), the effective potential changes very little within the critical horizon, approaching zero, as seen in the inset of Fig. \ref{p6}. Thus, photons move very slowly at this frequency, taking a long time to reach the inner light ring. The photons can be considered to be spread over the entire critical horizon, forming a light ball. This manifests as a region in the right panel, rather than a single ring. Additionally, it should be noted that the effective potential for the light ring at the origin is not meaningful, so photon motion at the origin is not discussed.
\section{CONCLUSION}\label{sec5}
In this paper, we construct Bardeen-Dirac Stars (BDSs) in AdS spacetime, which consist of two Dirac fields  and nonlinear electromagnetic field coupled with gravity. We investigate the properties of ADM mass, Noether charge and light rings of BDSs in AdS spacetime.

In AdS spacetime, the magnetic charge of BDSs cannot take arbitrary values like in Bardeen black holes but must be restricted within a critical magnetic charge. The solutions corresponding to this critical magnetic charge describe Bardeen black holes without an event horizon, meaning that the introduction of the Dirac field prevents the Bardeen model from having a black hole solution. This critical magnetic charge is not only constrained by the coupling constant $s$ but also influenced by the cosmological constant. A smaller cosmological constant results in a larger critical magnetic charge, thus extending the range of allowed solutions. Additionally, we observe that as the magnetic charge increases, the frequency of the BDSs approaches zero. When the frequency becomes extremely close to zero, BDSs exhibit a freezing phenomenon, which is referred to as Frozen Bardeen-Dirac Stars (FBDSs). FBDSs possess a critical horizon that encloses all the matter. When the cosmological constant is varied, the freezing behavior of BDSs disappears. Using the effective potential, we also study the light rings of BDSs and discover that they feature two light rings: a stable inner light ring and an unstable outer light ring. For smaller values of $q$, there exists a special frequency point where the two light rings merge into one. For the FBDSs, the inner light ring is actually a light ball where photons can move stably throughout the entire region. The cosmological constant affects the positions of the light rings, with a smaller cosmological constant causing the inner and outer light rings to move farther apart.

Our work extends the study of BDSs from asymptotically Minkowski spacetime to AdS spacetime. Our interest in BDSs extends beyond this work, as we aim to explore various extensions, such as rotating and charged BDSs in AdS spacetime. Additionally, the dynamical stability of BDSs is a promising area for further investigation.

\section*{Acknowledgements}
This work is supported by the National Key Research and Development Program of China (Grant No. 2020YFC2201503 and Grant No. 2021YFC2203003) and the National Natural Science Foundation of China (Grant No.~12047501 and No.~12275110).

\end{document}